\documentclass{jaa}
\usepackage{natbib}
\usepackage[colorlinks]{hyperref}
\usepackage{amsmath}
\usepackage{breqn}
\usepackage{psfrag}

\newcommand{\MSUN}{{\rm M}_{\odot}}

\usepackage[normalem]{ulem}

\usepackage{graphicx}

\begin{document}\sloppy

\title{Studying Cosmic Dawn using redshifted HI 21-cm signal: A brief review}


\author{Ankita Bera\textsuperscript{1,*}, Raghunath Ghara\textsuperscript{2}, Atrideb Chatterjee\textsuperscript{3}, Kanan K. Datta\textsuperscript{4}, Saumyadip Samui\textsuperscript{1,5}}
\affilOne{\textsuperscript{1}Department of Physics, Presidency University, 86/1 College Street, Kolkata 700073, India.\\}
\affilTwo{\textsuperscript{2}ARCO (Astrophysics Research Center), Department of Natural Sciences, The Open University of Israel, 1 University Road, PO Box 808, Ra’anana 4353701, Israel.\\}
\affilThree{\textsuperscript{3}National Centre for Radio Astrophysics, TIFR, Post Bag 3, Ganeshkhind, Pune 411007, India.\\}
\affilFour{\textsuperscript{4}Department of Physics and Relativity and Cosmology Research Centre, Jadavpur University, 188, Raja S.C. Mallick Rd, Kolkata 700032, India.\\}
\affilFive{\textsuperscript{5}School of Astrophysics, Presidency University, 86/1 College Street, Kolkata 700073, India.\\}


\twocolumn[{

\maketitle

\corres{ankita1.rs@presiuniv.ac.in}

\msinfo{XXX}{YYY}

\begin{abstract}
In this review article, we briefly outline our current understanding of the physics associated with the HI 21-cm signal from cosmic dawn. We discuss different phases of cosmic dawn  as the ambient gas and the background radiations evolve with the redshift. We address the consequences of several possible heating sources and radiation background on the global 21-cm signal. We further review our present perspective of other important aspects of the HI 21-cm signal such as the power spectrum and imaging. Finally, we highlight the future key measurements of the Square Kilometre Array and other ongoing/upcoming experiments that will enlighten our understanding of the early Universe.
\end{abstract}


}]


\doinum{12.3456/s78910-011-012-3}
\artcitid{\#\#\#\#}
\volnum{000}
\year{0000}
\pgrange{1--}
\setcounter{page}{1}
\lp{22}

\section{Introduction}
Our Universe came into existence with a Big Bang at approximately $13.6$ billion years ago. Universe got cooled as it expanded adiabatically, and over time, different ingredients of our Universe froze out. Quarks froze out first, then protons, and neutrons, followed by electrons. Finally, about $370,000$ years after the Big Bang, hydrogen, the first atoms, started to form by recombination of protons with the free electrons. With the formation of these first hydrogen atoms, the universe entered in a period called `Dark age'. During this time, hydrogen was mostly neutral until the first stars, quasars, and/or the first generation of galaxies appeared. The ignition of the first stars marked the end of the Dark age and the beginning of `Cosmic Dawn' (CD). It happened approximately $100$ million years after the Big Bang. However, it is extremely difficult to directly observe these early Universe faint sources as they are beyond the reach of our present-day telescopes. Therefore, observations of redshifted 21-cm signal from the neutral hydrogen have became the most promising tool for studying the cosmic dawn \citep{furlanetto06, pritchard12, 2019BAAS...51c..48C, 2021IJMPD..3030009P} and the subsequent epoch of reionization (EoR). 

There are mainly two separate approaches by which the cosmological HI 21-cm signal from the CD/EoR can be detected, namely, measurements of (i) global HI 21-cm signal, and, (ii) statistical signal such as HI power spectrum, Bi-spectrum etc. Several ongoing and upcoming experiments such as the Experiment to Detect the Global reionization Signature (EDGES)\citep{EDGES18}, Shaped Antenna measurement of the background RAdio Spectrum (SARAS)\citep{Saurabh_2021}, Large-Aperture Experiment to Detect the Dark Ages (LEDA) \citep{2018MNRAS.478.4193P}, the Radio Experiment for the Analysis of Cosmic Hydrogen (REACH) \citep{unknown} etc. are dedicated to detect the global HI 21-cm signal. However, the global HI 21-cm signal can not retain information regarding the spatial distribution of HI field and sources, and provide only the global evolution of the HI differential brightness temperature. Radio interferometric telescopes such as the Giant Metrewave Radio Telescope (GMRT) \citep{Ali_2008, Ghosh_2012, Pal21}, the Murchison Widefield Array (MWA) \citep{2016ApJ...833..102B, 2016MNRAS.460.4320E,patwa21},  Low Frequency Array (LOFAR) \citep{2017ApJ...838...65P,martens20}, the Hydrogen Epoch of Reionization Array (HERA) \citep{2017PASP..129d5001D, hera22}, and the Square Kilometre Array (SKA) \citep{2015aska.confE..10M} are devoted to probe the spatial fluctuations in the HI 21-cm signal during cosmic dawn and reionization. Along with the statistical signal the SKA should also be able to provide images of HI 21-cm field coming from cosmic dawn and EoR. 

In the view of these ongoing/upcoming facilities, it is extremely important that we have a thorough understanding of the cosmological HI signal expected from CD and EoR. There have been enormous progresses both on the theoretical understanding of the signal and observational side over the last two decades. Here, we present a short review on the cosmological HI 21~cm signal from cosmic dawn. In particular, we focus on the first luminous sources (e.g., Population~III and Population~II type stars), impact of various radiation backgrounds (e.g., Ly-$\alpha$, X-ray, Radio background) and physical processes (IGM heating due to cosmic rays, magnetic fields, X-ray ) on the HI 21-cm signal during the cosmic dawn. Readers are requested to follow the other companion review article \citet{skamo2}, which focuses on the HI 21-cm signal from the epoch of reionization and various observational challenges in order to detect the signal.

The outline of this article is as follows: we briefly outline the fundamentals of 21-cm signal in Section~\ref{sec:basic21cm} Section~\ref{th:sources} discusses our understanding on the early sources (Pop III and Pop II ) during cosmic dawn. Further, different radiation backgrounds, and
possible heating/cooling mechanisms of inter-galactic medium during cosmic dawn are described in Section~\ref{sec:rad_par} and \ref{sec:heat_source} respectively. Section~\ref{sec:global} focuses on global HI 21~cm signal resulting from various semi-analytical models and simulations.  Next, we focus on the current status of the understanding of HI 21-cm signal power spectrum (Section~\ref{sec:ps}). Then in Section~\ref{param_est}, we discuss prospects of parameter estimation using measured HI power spectrum. Further, the topological aspects of the 21~cm signal is presented in Section~\ref{sec:image}
Subsequently in Section~\ref{sec:synergy}, we discuss the prospects of a joint analysis of the statistical and global signal for better constraining the models of CD. Finally, in Section~\ref{sec:summary}, we present a summary. Throughout this article we assume a flat, $\Lambda$CDM cosmology  with the cosmological parameters obtained from recent Planck 2018 \citep{Planck18} observation, i.e. $\Omega_{\Lambda} = 0.69$, $\Omega_{\rm m} = 0.31$, $\Omega_{\rm b} = 0.049$, and the Hubble parameter $H_0 = 67.66$~km/s/Mpc, unless otherwise mentioned.

\section{Fundamentals of 21 cm signal: Global signal and power spectrum}
\label{sec:basic21cm}
The quantity of interest in the context of the cosmological 21-cm signal is the differential HI brightness temperature ($\delta T_{b}$). It is defined as the excess brightness temperature relative to a background radio temperature and redshifted to the present observer, and is given by \citep{Bharadwaj_2005},
\begin{multline}
    \delta T_{b}(\textbf{n},z) = 4.0 \,{\rm mK} (1+z)^2 \left( \frac{\Omega_b h^2}{0.02} \right) \left(\frac{0.7}{h} \right) \left(\frac{H_0}{H(z)} \right) \left(\frac{\rho_{\rm HI}}{\Bar{\rho_H}} \right) \\ \left( 1-\frac{T_{\gamma}}{T_s} \right) \left[1- \frac{(1+z)}{H(z)} \frac{\partial v}{\partial r} \right] .
    \label{eq:T21_full}
\end{multline}
Here, $\rho_{\rm HI}$ is the density of the neutral hydrogen whereas $\Bar{\rho}_{\rm H}$ is the mean hydrogen density, and $\textbf{n}$ is the direction of light propagation. 
Further, $({\rho_{\rm HI}}/{\Bar{\rho}_H})$ arises due to the non-uniform distribution of hydrogen, and the term inside the square brackets arises due to the redshift space distortion in which $\partial v/ \partial r$ is the divergence of the peculiar velocity along the line of sight \citep{Bharadwaj_2004}. Moreover, $T_{\gamma}$ is the background temperature of radio photons, mostly dominated by the cosmic microwave background radiation (CMBR) but there could be other candidates that can produce background radiation (we discuss this in Sec.~\ref{th:radio_bkg}). $T_s$ is the hydrogen spin temperature which is determined by the relative population of the singlet and triplet states of neutral hydrogen atom.
It is clear from Eq.~\ref{eq:T21_full} that 21-cm signal will be in the absorption or emission depending on whether $T_s<T_{\gamma}$ or $T_s>T_{\gamma}$ respectively.
The spin temperature is related to the gas kinetic temperature $T_{\rm K}$ and background temperature $T_{\gamma}$ as (\citealt*{Field_1958}; also see \citealt*{furlanetto06} for a detailed review), 
\begin{equation}\label{spin_temp}
    T^{-1}_s = \frac{T^{-1}_{\gamma} + x_{\alpha} T^{-1}_{\alpha} + x_c T^{-1}_g}{1 + x_{\alpha} + x_c} ,
\end{equation}
where, $T_{\alpha}$ is the color temperature corresponds to the Lyman-$\alpha$ radiation field. As the Lyman-$\alpha$ photons get absorbed and emitted repeatedly by hydrogen atoms, they are in equilibrium with H-atom, so $T_{\alpha} = T_{\rm K}$ during cosmic dawn period. The coupling coefficients, $x_c$ and $x_{\alpha}$ depend on the different processes such as Ly-$\alpha$ coupling \citep[due to Wouthuysen-Field mechanism][]{Wouthuysen_1952, Field_1958}, and collisional coupling due to the collisions between two hydrogen atoms, hydrogen atom and an electron or the H-atom and a proton.

The Wouthuysen-Field coupling coefficient is given by \citep{pritchard12},
\begin{equation}
    x_{\alpha} = \frac{16 \pi^2 T_{*} e^2 f_{\alpha}}{27 A_{10} T_{\gamma} m_e c} S_{\alpha} J_{\alpha} ,
    \label{x_a}
\end{equation}
where $f_{\alpha} = 0.4162$ is the oscillator strength for the Ly-$\alpha$ transition. Further, $J_{\alpha}$ is the Ly-$\alpha$ photon intensity which will be discussed in Sec.~\ref{th:ly-alpha}. Moreover, $S_{\alpha}$ is a correction factor of order unity which takes care of the redistribution of photon energies due to the repeated scattering off the thermal distribution of atoms \citep{Chen_2004}. Also, $T_* = h_p \nu_e/k_B = 0.068$~K is the characteristic temperature for the HI 21-cm transition.
The total collisional coupling coefficient can be written as a sum of coupling between H-H, H-p, H-$e^-$, ($x_c^{\rm HH}, x_c^{\rm pH}, x_c^{\rm eH}$ respectively), and is given by,
\begin{eqnarray}
    x_c & = & x_c^{\rm HH} + x_c^{\rm eH} + x_c^{\rm pH}   \\
    & = & \frac{T_*}{A_{10} T_{\gamma}} {\kappa^{\rm HH}_{10}(T_{\rm K}) n_{\rm H} + \kappa^{\rm eH}_{10}(T_{\rm K}) n_{\rm e} + \kappa^{\rm pH}_{10}(T_{\rm K}) n_{\rm p}} . \nonumber
    \label{x_c}
\end{eqnarray}
All the specific rate coefficient values, $\kappa^{\rm HH}_{10}, \kappa^{\rm eH}_{10}$, and $\kappa^{\rm pH}_{10}$ are given in \citet{pritchard12}. As the Universe was mostly filled with hydrogen during cosmic dawn, $\kappa^{\rm HH}_{10}$ dominate over $\kappa^{\rm eH}_{10}$, and $\kappa^{\rm pH}_{10}$ throughout this period.
Although, this collisional coupling is a dominant process during dark ages, so the coupling between $T_s$ and $T_{\rm K}$ happens due to the Ly-${\alpha}$ coupling during cosmic dawn.

Note that, $\rho_{\rm HI}$, and the number density of hydrogen ($n_{\rm H}$), electron ($n_{\rm e}$), and proton ($n_{\rm p}$) are determined by the ionization state of the inter-galactic medium (IGM). This can be obtained by the evolution in the ionized fraction of hydrogen ($x_e$) which can be written as,
\begin{dmath}
    {\frac{dx_e}{dz}=} \left[ C_p \left( \beta_e(T_{\gamma})\,(1-x_e)  \, e^{-\dfrac{h_p\nu_\alpha}{k_B T_{\rm K}}} - \alpha_e(T_{\rm K})\,x_e^2 n_{\rm H}(z) \right) \\   + \gamma_e(T_{\rm K})\,n_{\rm H}(z) (1-x_e)x_e + \frac{\dot N_{\gamma}}{n_{\rm H}(z)} \right] \frac{dt}{dz}.
    \label{eq:ion_frac}
\end{dmath}
Here the evolution in ionization fraction is affected due to the photoionization by CMBR photons, recombination, collisional ionization, and photoionization by UV photons respectively. The photoionization co-efficient, $\beta_e$ can be calculated using the relation, $\beta_e (T_{\gamma}) = \alpha_e (T_{\gamma}) \Big( \frac{2 \pi m_e k_B T_\gamma}{h^2_p} \Big)^{3/2} e^{-E_{2s}/k_B T_\gamma}$ \citep{Seager1999,Seager2000}. The recombination co-efficient, $\alpha_e (T_{\rm K}) = F \times 10^{-19} (\frac{a t^b}{1 + c t^d}) \hspace{0.1cm} {\rm m^3 s^{-1} } $, where $a = 4.309$, $b = -0.6166$, $c = 0.6703$, $d = 0.53$, $F = 1.14$ (the fudge factor)  and $t = \frac{T_{\rm K}}{10^4 \, {\rm K}} $. The Peebles factor is defined by $C_p = \frac{1+ K \Lambda (1-x) n_{H}}{1+K(\Lambda+\beta_{e})(1-x) n_{H}}$, where $\Lambda=8.3 \, {\rm s}^{-1}$ is the transition rate from (hydrogen ground state) $2s\rightarrow1s$ state through two photons decay, and $K=\frac{\lambda_{\alpha}^{3}}{8\pi H(z)}$. The collisional ionization coefficient, $\gamma_e(T_{\rm K}) =0.291 \times 10^{-7} \times U ^{0.39} \frac{\exp(-U)}{0.232+U}  \, {\rm cm^3/s} $ \citep{Minoda17} with $h_{p}  \nu_{\alpha} = 10.2 \, {\rm eV}$ and $U=\vert E_{1s}/k_B T_{\rm K} \vert$. Further, $\dot N_{\gamma}$ is the rate of UV photons escaping into the IGM and $n_{\rm H}(z)$ is the proper number density of the hydrogen atoms \citep{barkana01}.

Global signal experiments attempt to detect the sky-averaged $\langle \delta T_{b} \rangle$ where the average is taken over all directions of sky at a particular redshift. Thus the globally averaged differential brightness temperature, $\delta T_{b}(z)$ is
given by,
\begin{eqnarray}
    \delta T_{b}(z) =  4.0 \,{\rm mK} (1+z)^2 \left( \frac{\Omega_{\rm b} h^2}{0.02} \right) \left(\frac{0.7}{h} \right) \left(\frac{H_0}{H(z)} \right) \nonumber \\ \times \left(\frac{\rho_{\rm HI}}{\Bar{\rho_H}} \right) 
    \left( 1-\frac{T_{\gamma}}{T_s} \right).
    \label{eq:T21_global}
\end{eqnarray}
Experiments such as the EDGES, SARAS, LEDA, REACH etc. are trying to detect this global 21-cm signal. 

While single antenna based experiments can only measure the redshift evolution of the sky averaged HI 21-cm signal, the radio interferometers such as the GMRT, MWA, LOFAR, HERA are sensitive to the spatial fluctuations of the signal. Due to limited sensitivity,  the presently operating radio interferometers aim to measure these spatial fluctuations in terms of different statistical quantities such as the variance, power spectrum of the signal, etc \citep{Ali_2008, harker12, patil2014, martens20, 2021MNRAS.506.3717R}. The most straightforward way to measure these fluctuations in a radio interferometric observation is through the power spectrum, which is the Fourier transform of the 2-point correlation function of the signal. The power spectrum of the HI 21-cm signal $P_{21}(k,z)$ can be expressed as,
\begin{equation}
    \langle \Tilde{T_b}(\textbf{k},z) \Tilde{T^*_b}(\textbf{k}^{'},z) \rangle = (2\pi)^3 \delta^D (\textbf{k} - \textbf{k}^{'}) P_{21}(k,z) ,
\end{equation}
where, $\Tilde{T_b}(\textbf{k},z)$ is the Fourier transform of $\delta T_{b}(\textbf{n},z)$, $\delta^D$ is the 3D dirac delta function while $\textbf{k}$ and ${\textbf{k}^{'}}$ are the comoving wavevectors.
Normally the power spectrum is represented in terms of the dimensionless quantity as,
\begin{equation}
    \Delta^2(\textbf{k},z) = \frac{k^3 P_{21}(\textbf{k},z)}{2\pi^2} .
    \label{eq:dimps}
\end{equation}
This quantity also represents the power per unit logarithmic interval in $k$-scale. The redshift space HI 21-cm power spectrum carries information about the large scale distribution of the HI field, dark matter and, thus, can be decomposed into different components related to power spectra of pure dark matter, HI fraction etc \citep{mao12, majumdar13, ghara15a}.

Although the currently operating radio interferometers have limited sensitivity,  upcoming low-frequency telescope such as SKA1-low will have $\sim 10$ times higher sensitivity than the telescope like LOFAR. With such a high sensitivity, SKA1-low will be able to detect the epoch of reionization and cosmic dawn signal statistically, and to produce the tomographic images of the HI 21-cm signal from these epochs as well. We will discuss prospects of detecting such HI  images later in section \ref{sec:image}. These tomographic images, as well as $\delta T_{b}(z)$ and $\Delta^2(k,z)$ of the signal from the cosmic dawn and EoR, are crucially dependent on the properties of the sources present during these epochs. The source dependence in Eq.~\ref{eq:T21_full} appears through the neutral fraction of hydrogen ($\frac{\rho_{\rm HI}}{\Bar{\rho_H}}$), spin temperature ($T_S$) and any excess radio background produced from the sources. We shall discuss how different types of sources and radiation backgrounds impact the signal in the following sections.

\section{Early sources during cosmic dawn}
\label{th:sources}
After the epoch of recombination, for the first few hundred million years, the Universe was mostly filled with neutral hydrogen and helium atoms.
In the hierarchical structure formation, the dark matter halos were formed by gravitational collapse. The baryons (primordial hydrogen and helium) were pulled into the potential wells created by the dark matter halos. As the gas cools and gas mass exceeds the Jeans mass, first stars started to form. Existing theoretical studies \citep{bromm04,abel02} suggest that the first stars thus formed were massive, luminous, and metal-free, known as Population~III (hereafter Pop~III) stars. They are likely to produce copious amount of UV photons, and strongly affect the high redshift IGM, in turn the cosmic 21-cm signal during cosmic dawn \citep{fialkov14, 2015MNRAS.448..654Y, 2018MNRAS.478.5591M, Mebane_2018, 2018MNRAS.480.1925T, 2019ApJ...877L...5S, atri20, Bera2022,  2022MNRAS.511.3657M, 2022arXiv220102638H}. 

The first stars are likely to form in minihalos with virial temperature $T_{\rm vir} \sim 300-10^4$ K that can cool via molecular $H_2$ cooling \citep{barkana01}.
The $H_2$ cooling depends on the amount of molecular hydrogen present in a halo, which can be dissociated in presence of a background of Lyman-Werner (LW) photons (see details in Sec.~\ref{th:LW}). Thus, whether a halo can host Pop~III stars or not depends on the critical amount of molecular hydrogen present along with the feedback of LW radiation created by the first stars themselves \citep[see][for detailed modelling of Pop~III star formation]{Tegmark_1997, Mebane_2018, Bera2022}.
It was shown by \citet{Tegmark_1997} that the $H_2$ fraction of a halo varies as,
\begin{equation}
    f_{\rm H_2} \approx 3.5 \times 10^{-4} T_3^{1.52}
    \label{eq:fH2}
\end{equation}
where $T_3 = T_{\rm vir}/10^{3}$~K, and a halo can host Pop~III stars if this fraction exceeds a critical value, given by,
\begin{eqnarray}
    f_{\rm crit,H_2} \approx 1.6 \times 10^{-4} \left( \frac{1+z}{20} \right)^{-3/2} \left( 1+ \frac{10 T_3^{7/2}}{60+T_3^4} \right)^{-1}  \nonumber \\
     \exp\left( \frac{0.512 K}{T_3} \right) .
    \label{eq:f_cH2}
\end{eqnarray}
Note that, the star formation mechanism in minihalos are still ongoing research topic, and in particular the initial mass function of Pop~III stars are poorly constrained \citep{abe2021,parsons2021,lazar2022, Gessey2022}.
Several people use the top heavy IMF model (mass of star, $M_* >100 M_{\odot}$) due to metal-free cooling and inefficient fragmentation \citep{bromm99,abel02}. Even a halo can host a single such massive star. Further, the final stages of these stars, and their effect on IGM are also governed by their initial masses.
In fact, the HI 21-cm signal from cosmic dawn would provide a very good test-bed for these models.

Once the first generation of stars enrich the IGM with metals, the metal-enriched Population~II (hereafter Pop~II) stars began to form in the atomic cooling halos. The properties of these stars are quite similar to the stars that we see today, and their properties are much more constrained compared to Pop~III stars. There already exist several models of Pop~II stars that are well-constrained by different observational evidence \citep{Galform, 2005MNRAS.361..577C, 2006MNRAS.371L..55C, Galacticus,  Samui_2014,  dayal2018}.  
The number density of the pop III stars is expected to decrease rapidly at $z\lesssim 15$ once the LW feedback becomes significant. Thus, the major contribution to the ionization of the IGM neutral hydrogen is expected to come from Pop II stars inside the first galaxies. Given a dark matter halo of mass $M_{\rm halo}$, the amount of stellar mass contained, intrinsic spectral energy distribution, IMF, escape fraction of these ionizing photons, etc. are still uncertain for these early galaxies. Thus, numerical simulations of EoR generally work under simplified pictures of Pop II star formation. Most of these simulations assume simple scaling relations between the total stellar mass  and the hosting dark matter halo mass \citep{ghara15b, 2019MNRAS.487.1101R, 2021MNRAS.501....1G}. The impact of the fraction of baryons residing within the stars in a galaxy, the average number of ionizing photons per baryon produced in the stars,  and the escape fraction of the UV photons are degenerate on the ionization/thermal states of the IGM. Thus, many studies ( e.g., \citet{iliev12, ghara2020, 2022MNRAS.511.2239M}) treat the product of all these quantities as a single parameter termed as ionization efficiency parameter. While semi-numerical simulations (e.g., \citet{majumdar11, 2021MNRAS.501....1G, 2022MNRAS.511.2239M}) do not have provision to adopt the spectral energy distributions (SEDs) of the galaxies, the radiative transfer simulations such as \citet{mellema06, 2011MNRAS.414..428P, ghara15b} use SEDs of such early galaxies assuming simple forms such as a blackbody spectrum or using a SED generated from population synthesis codes such as PEGASE \citep{Fioc97}. The ionizing photons emitted from the galaxies not only ionize the neutral hydrogen in the IGM but also suppress star formation in low-mass galaxies due to thermal feedback. Thus, the star formation inside dark matter halos with mass $M_{\rm halo} \lesssim 10^9 ~M_{\odot}$ becomes inefficient if those halos remained inside ionized regions \citep{ghara18, 2021MNRAS.506.3717R}.

Apart from the stars and galaxies, impact of  mini-quasars and quasars have also been studied in the context of HI 21-cm signal from cosmic dawn \citep{2001ApJ...563....1V, thomas11, ghara15a}.
As soon as the first luminous sources appeared, they started to emit UV photons, Lyman-series photons, and X-rays. X-rays may also be produced by different sources such as supernova remnants (SNR), and miniquasars \citep{2003MNRAS.340..210G, 2006MNRAS.371..867F,2016ASSL..423....1H}. Some contribution of UV photons are also expected from quasars and AGNs \citep{2010AJ....140..546W,2015ApJ...813L...8M}. All these are likely to produce feedback on IGM as well as the subsequent star formation. Thus one needs to model these in a self consistent manner, and we describe them in the next section.

\section{Radiation background}
\label{sec:rad_par}

\subsection{Ly-$\alpha$ background} 
\label{th:ly-alpha}
With the formation of first generation of stars, the 
photons get emitted at all frequencies depending on the mass of the stars, and they affect the surrounding medium. In particular, the Ly-$\alpha$ photons alter the spin states of the neutral hydrogen atom, and thus affect the spin temperature, $T_s$. Therefore one needs to calculate the Ly-$\alpha$ background radiation generated from first stars to obtain the 21-cm signal.
A photon having frequency greater than Ly-$\alpha$ gets redshifted into a Ly-$n$ series photon due to the expansion of the Universe. Such a photon can get absorbed by the ground state hydrogen atom, and then there are two possibilities through which Ly-$\alpha$ photon gets generated. Either this excited hydrogen atom cascades down to $n=2$ state first and finally from $n=2$ to $n=1$ state by producing a Ly-$\alpha$ photon, or directly jumps into the ground state which produces another Ly-$n$ photon that can eventually produce a Ly-$\alpha$ photon by the first process.
Readers are requested to check the details on the cascading of Ly-$n$ photons given in \citet{Pritchard_2006}.

The Ly-$\alpha$ background is generally estimated using proper Ly-$\alpha$ photon intensity which is defined as the spherically averaged number of photons striking per unit area, per unit frequency, per unit time, and per steradian. It is given by,
\begin{equation}
    J_{\alpha} = \frac{(1+z)^2}{4 \pi} \sum_{n=2}^{n_{\rm max}} f_{\rm recycle}(n) \int_{z}^{z_{\rm max}(n)} \frac{c dz^{'}}{H(z^{'})} \epsilon(\nu_n^{'},z^{'}) ,
    \label{eq:J_alpha}
\end{equation}
where, the summation over the atomic level $n$ is truncated at $n_{max} \simeq 23$ to exclude the levels for which the horizon resides within HII region of an isolated galaxy as pointed out by \citet{Pritchard_2006}. Further, the probability of generating a Ly-$\alpha$ photon from a Ly-$n$ photon is characterised by $f_{\rm recycle}(n)$, and the values for different levels can be obtained from \citet{Pritchard_2006}. As the cascading of Ly-$n$ photons actually delays the onset of strong Wouthuysen-Field coupling, one should include that in the Ly-$\alpha$ flux calculation. Moreover, The absorption at a level $n$ at redshift $z$ corresponds to a frequency which is emitted at a higher redshift, $z^{'}$, and this frequency can be written in terms of Lyman limit frequency, $\nu_{\rm LL}$ as \citep{Barkana_2005},
\begin{equation}
    \nu_n^{'} = \nu_{\rm LL} (1-n^{-2}) \frac{1+z^{'}}{1+z} .
\end{equation}
But the photon to be available at Ly-$\alpha$ resonance at a redshift $z$, should have been emitted below a redshift of,
\begin{equation}
    z_{\rm max}(n) = (1 + z) \frac{[1 - (1+n)^{-2}]}{(1-n^{-2})} - 1,
\end{equation}
so that it can participate in this process.
In this way, the contribution of photons emitted between consecutive atomic levels to the total flux get summed up. Moreover, we multiply $J_{\alpha}$ with the normalization parameter $f_{\alpha}$ to incorporate the uncertainty in the IMF of the first stars and escape of Ly-$\alpha$ photons. Further, $\epsilon(\nu, z)$ in Eq.~\ref{eq:J_alpha} is the comoving photon emissivity, defined as the number of photons emitted by stars at redshift $z$ per unit comoving volume, per proper time and frequency, and at rest frame frequency $\nu$. It is generally obtained from the star formation rate density, $\dot \rho_*$ at a given redshift, and is related by,
\begin{equation}
    \epsilon(\nu, z) = \frac{\dot \rho_*}{m_p} \epsilon_b(\nu) ,
\end{equation}
where, $m_p$ is the mass of proton. The spectral distribution function of the sources, or the number of photons produced per baryon of stars, $\epsilon_b(\nu)$ depends on the initial masses and the composition of the stars. It is generally modelled as a power law $\epsilon_b(\nu) \propto \nu^{\alpha_s - 1}$ with the spectral index of $\alpha_s$. The values of $\alpha_s$ for Pop~III and Pop~II stars are $1.29$ and $0.14$ respectively. The spectral distribution function is normalised to emit $4800$ and $9690$ photons per baryon between Ly-$\alpha$ and Lyman limit frequencies for Pop~III and Pop~II stars respectively, whereas the corresponding numbers between Ly-$\alpha$ and Ly-$\beta$ frequencies are $2670$ and $6520$ \citep{Barkana_2005}. 
Note that, apart from the coupling of $T_s$ to $T_{\rm K}$, Ly-$\alpha$  could also heat up the IGM which has been studied in \citet{Chen_2004,Furlanetto06MNRAS,ghara2020,mittal21}.

\subsection{Lyman-Werner background}
\label{th:LW}
The first generation of stars not only emits the Ly-$\alpha$ photons but also produces a background of Lyman-Werner (LW) radiation. The photons in the LW band possess energy in the range between $11.5$-$13.6$ eV, and can photo-dissociate the molecular $H_2$. 
As the Pop~III stars form in a halo where gas cools via molecular hydrogen cooling mechanism, the presence of LW photons acts as a negative feedback for Pop~III star formation. If a halo is present in a LW background, the Pop~III star formation gets affected depending on the amount of LW flux and the feedback can even completely stop Pop~III star formation in low mass halos. Therefore the required minimum mass for a Pop~III star to form as mentioned in Sec.~\ref{th:sources} gets modified \citep[for deatils see][]{Mebane_2018}. 

The flux of Lyman-Werner background can be written as \citep{visbal14},
\begin{equation}
    J_{\rm LW}(z) = \frac{c}{4 \pi} \int_z^{z_m} \frac{dt}{dz^{'}} (1+z)^3 \epsilon(z^{'}) dz^{'} ,
    \label{eq:J_LW}
\end{equation}
where, $c$ is the speed of light, and $z_m$ is the maximum redshift that a photon gets generated and redshifted into the Lyman series at redshift $z$. It can be calculated using the relation, $\frac{1+z_m}{1+z} = 1.04$, assuming a $4 \%$ of photons in the LW band get redshifted before hitting Ly-$\alpha$ line \citep{visbal14}.
Further, $\epsilon(z^{'})$ is the specific LW comoving luminosity density which again can be obtained from the star formation rate density as,
\begin{equation}
    \epsilon(z) = \frac{\dot \rho_{*}}{m_p} \left( \frac{N_{\rm LW} E_{\rm LW}}{\Delta \nu_{\rm LW}} \right) .
    \label{eq:eps_z}
\end{equation}
Here, the average energy of a LW photon, $E_{\rm LW} = 11.9\,{\rm eV}$, and the LW frequency band, $\nu_{\rm LW} = 5.8 \times 10^{14}$ Hz \citep{Mebane_2018}. Further, $N_{\rm LW}$ is the number of photons per baryon of stars in the energy range $11.5-13.6$~eV, and the corresponding numbers for Pop~III and Pop~II stars are already mentioned in the previous Sec.~\ref{th:ly-alpha}.

\subsection{Excess radio background}  
\label{th:radio_bkg}
In the standard galaxy formation modelling, the background radiation $T_{\gamma}$ (appeared in Eq. \ref{spin_temp} ) is assumed to have contribution only from CMB radiation. However, observation with LWA1 \citep{2018ApJ...858L...9D}, ARCADE-2 \citep{2011ApJ...734....5F} reported detection of excess radio background. The origin of this excess radio background is still not clear. For example, \cite{2011ApJ...734....6S} argued that our own Milky Way galaxy can produce this radio background, whereas some other studies  \citep{mirocha2019, 2018ApJ...858L..17F, 2020MNRAS.492.6086E, 2022MNRAS.510.4992M} have proposed that high-$z$ sources like radio loud black holes, bright luminous galaxies, Pop~III supernova and even primordial black holes \citep{2022MNRAS.510.4992M} can produce such excess radio background. If we assume the later scenario to be true, then it becomes necessary to model the background temperature $T_{\gamma}$ to have contribution from both CMBR and the excess radio background, and therefore, $T_{\gamma}$ should be expressed as
\begin{equation}
    T_{\gamma}= T_{\rm CMB} + T_{R} ,
\end{equation}
where, $T_{\rm CMB}$ is the CMBR temperature and $T_{R}$ is the temperature coming from the excess radio background.

As shown in \citet{atri20}, the radiation flux received at redshift $z$, can be computed using
\begin{equation}
F_{R}(z)=  \left(\frac{1420}{150}\right)^{\alpha_R} \frac{c(1+z)^{3}}{4\pi} \int_z^{\infty} \epsilon_{R, \nu'}(z') \left|\frac{d t'}{d z'} \right| d z',
\end{equation}
where  $\epsilon_{R, \nu'}(z')$ is the comoving radio emissivity at $\nu' = 150~{\rm MHz} (1 + z') / (1 + z)$, and $\alpha_{R}$ is the radio spectral index which is usually assumed to be $-0.7$ \citep{gurkan2018}. Once the flux is computed, one can then calculate the radio background temperature $T_{R}$ using the Rayleigh-Jeans law.

\section{Possible heating/cooling sources during cosmic dawn}
\label{sec:heat_source}
The IGM kinetic temperature is an important quantity that needs to be estimated for calculating the spin temperature (see Eq.~\ref{spin_temp}) and the HI 21-cm signal during cosmic dawn. The evolution of the IGM temperature $T_K$ can be written as,
\begin{multline}
    \frac{dT_{\rm K}}{dz} = \frac{2T_{\rm K}}{1+z} - \frac{8\sigma_T a_{\rm SB} T_{\gamma}^4}{3m_e c H(z) (1+z)}\left(T_\gamma-T_{\rm K}\right) \frac{x_e}{1+x_e}  \\
    - \frac{2}{3 n_H k_B} \sum_i Q_i ,
	\label{eq:Tg}
\end{multline}
where, $ k_{\rm B}$, $ \sigma_{\rm T}$, and $\sigma_{\rm SB}$ are the Boltzmann constant, Thomson scattering cross-section, and Stefan Boltzmann constant respectively. In the first two terms, here we consider the adiabatic cooling due to the expansion of the Universe and the Compton heating due to the interaction between CMBR and free electrons. Further, $Q_i$ includes any other possible heating/cooling processes that we describe next.
There are several possible mechanisms by which the IGM can be heated up such as heating due to soft X-ray, cosmic rays, Ly-$\alpha$ photons, magnetic field, Dark Matter decay/annihilation \citep{pritchard2007, ghara15a, ghara2020, sethi05, Sazonov_2015, Chen_2004, SS05, furlanetto2006, liu2018}, and we discuss some of them in the following sections.

\subsection{X-ray Heating}
\label{th:x-ray}
X-ray photons can be produced by early sources such as, accreting black holes, miniquasars, supernova shocks or X-ray binaries, that can easily escape from galaxies, and can increase the temperature of the IGM. The determining factors for the X-ray heating are the number and the spectral shape of the X-ray photons. As no observational evidence exists in case of high redshift Universe, the X-ray heating modelling are limited by the uncertainty of both the aforementioned factors. Majority of the works, present in the literature, takes a conservative and simple approach \citep[e.g.,][]{furlanetto04c, furlanetto06, Mirocha_2019, 2021MNRAS.507.2405C}. This approach simply assumes a correlation between the star formation rate ($\dot{M}_*$) and the X-ray luminosity ($L_X$) of a galaxy at high-$z$, which is motivated by the observed correlation prevailing in the local Universe. It also includes a free parameter, $f_X$, which is an unknown normalization factor allowing one to take into account the differences between the local and high-$z$ Universe. The correlation can be expressed as \citep{2012MNRAS.419.2095M},
\begin{equation}
    L_X = 3.4 \times 10^{33} f_X \left(\frac{\dot{M}_*}{\mathrm{M}_{\odot}~\mathrm{yr}^{-1}} \right) \mathrm{J~s}^{-1}.
    \label{eq:Lx}
\end{equation}
The globally averaged X-ray energy density which helps in heating IGM can then be written as,
\begin{equation}
  Q_{X} =3.4 \times 10^{33} f_X \times f_{h} \left( \frac{\dot{\rho}_*}{\mathrm{M}_{\odot}~\mathrm{yr}^{-1}~\mathrm{Mpc}^{-3}} \right) {\mathrm{J ~s^{-1} Mpc^{-3}}},
\end{equation}
where, $\dot{\rho_{*}}$ represents the global star formation rate density whereas, $f_{h}$ denotes the fraction of X-ray which helps in heating the IGM. This value is generally taken to be $\sim 0.3$ for standard X-ray heating analysis \citep{furlanetto06}.

\subsection{Heating due to cosmic rays} 
\label{th:CRheat}
Along with X-rays, there are other possible candidates that can heat the IGM substantially. One such candidate is cosmic ray protons that are generated in the termination shocks of supernova explosion \citep{Sazonov_2015, Leite_2017, Jana_2019, Bera2022}.
It is well known that the SNe are the main sources for high energy cosmic rays. Given the expected top-heavy IMF of Pop~III stars, they are all likely to explode as a SNe that can accelerate copious amount of cosmic rays. Due to the smaller sizes of the host galaxies as well as the higher energetics of the Pop~III SNe, the cosmic rays are likely to be generated outside the virial radius of the halo and can escape to the IGM easily \citep{Sazonov_2015}. These cosmic rays (mostly the low energy protons) interact with the IGM via collision and transfer their energy to the thermal gas.

While traversing through IGM, the low energy protons ($\lesssim 30$ MeV) interact with the neutral hydrogen and free $e^-$. In case of collision with free $e^-$, the entire energy loss by these particles becomes the thermal energy of the IGM, whereas, the interaction between cosmic ray protons and neutral hydrogen results in the primary and secondary ionizations, and finally contribute to the heating \citep[see][for detailed analysis of cosmic ray interactions]{schlickeiser2002cosmic}. Thus cosmic rays resulting from Pop~III stars can potentially alter the thermal state of the IGM, and hence the spin temperature, $T_s$ during cosmic dawn \citep{Bera2022}.

On contrary to Pop~III halos, the low energy protons generated from the SNe exploding in massive atomic cooling halos, get confined within the galaxy itself. However, the high energy protons are likely to escape from these halos like our milky way galaxy, and can contribute to the heating. In this case, if a sufficient magnetic field is present in the IGM, these high energy cosmic ray particles gyrate along the magnetic field lines and excite magnetosonic Alfv\`en waves. When these waves get damped, the energy gets transferred to the thermal gas, and can potentially change the temperature of the IGM \citep{Kulsrud_1969, Skilling1975, Bell1978, Kulsrud_2004}.

In both cases of cosmic rays generated from Pop~III and Pop~II stars, the amount of heating depends on the total energy density of cosmic rays, and hence the efficiency with which the cosmic rays are accelerated in SNe. This efficiency, $\epsilon_{\rm CR}$ can be as high as $30 \%$ \citep{2005ApJ...620...44K}. The total energy density of cosmic rays can also be calculated from the star formation rate densities as,
\begin{eqnarray}
    \dot E_{\rm CR}(z) = 10^{-30} \epsilon_{\rm CR} \, f_{\rm SN} (1+z)^3 \left( \frac{E_{\rm SN}}{10^{51} \, erg} \right) \nonumber \\
    \times \left( \frac{\dot \rho_*(z)}{\rm M_{\odot}\, yr^{-1} \, Mpc^{-3}} \right)  ,
    \label{eq:E_CR}
\end{eqnarray}
where, $E_{\rm CR}(z)$ is the energy density per unit physical volume, and in units of ${\rm erg\,s^{-1} cm^{-3}}$.
Here, $E_{\rm SN}$ is the kinetic energy of the supernova which may differ for different type of supernovae such as, pair-instability SNe and core-collapse SNe that one considers. Further, the number of SNe explosions per solar mass of stars is $f_{\rm SN}$ which also depends on the initial mass function of the stars.

As mentioned earlier, in case of cosmic rays generated in Pop~II stars, the possible mechanism of heating is via the generation of magnetosonic waves which depends on the amount of magnetic field present in the IGM. Apart from this, the IGM magnetic field can also alter the temperature via ambipolar diffusion and decaying turbulence mechanisms that we discuss in the next section.

\subsection{Heating due to the magnetic field}
\label{th:PMF}
At the recombination period of the Universe, the primeval plasma recombines to form neutral hydrogen, and the ionization fraction decreases and finally reaches to $10^{-4}$ at redshift $100$. But the residual free $e^-$ is still sufficient to carry the current to sustain primordial magnetic field that may have generated in the very early Universe during inflation \citep[see][for reviews]{PhysRevD.37.2743, Ratra1992, Grasso2001, Giovannini2004}. This magnetic field exert forces on the $e^-$-ion fluids which causes a relative drift velocity, and consequently, a frictional force arises between the neutral and ionized components. This leads to the dissipation of magnetic energy, termed as ambipolar diffusion mechanism, and can heat the IGM. On the other hand, due to the recombination, the photon mean free path increases which cause a reduction in viscosity of the fluids. As a consequence, the fluids Reynolds number increases which generate the decaying fluid turbulence. This turbulence can transfer energy from the magnetic field to the IGM.
So, the magnetic field energy gets transferred to the IGM through the ambipolar diffusion and decaying magneto-hydrodynamic turbulence processes which is likely to the change the IGM temperature \citep{sethi05}.

A detailed modelling of the magnetic energy dissipation, and consequently the heating and ionization due to presence of magnetic field is discussed in \citet{sethi05, Minoda19, Bera_2020, bhatt2020, Natwariya2020}.
The heating rate (in units of energy per unit time per unit volume) due to the ambipolar diffusion, $\Gamma_{\rm AD}$ can be written as,
\begin{eqnarray}
    \Gamma_{\rm AD} = \frac{(1-x_e)}{\gamma x_e \rho_b^2} \frac{\Big \langle \vert (\nabla\times \boldsymbol{B})\times \boldsymbol{B} \vert^2 \Big \rangle}{16\pi^2},
    \label{eq:gamma_AD}
\end{eqnarray}
where, $\rho_b$ is the baryon mass density at redshift $z$, and the coupling coefficient between the ionized and neutral components is $\gamma= 1.94 \times 10^{14} \, (T_{\rm K}/{\rm K})^{0.375} \, {\rm cm}^3 {\rm gm}^{-1} {\rm s}^{-1}$ \citep{sethi05, Chluba15}. Here, the Lorentz force is approximated as, $ \Big \langle \vert (\nabla\times \boldsymbol{B})\times \boldsymbol{B} \vert^2 \Big \rangle$, and the detailed calculation of it is given in \citet{Chluba15, KK14}.

Another heating rate due to the decaying turbulence is given by,
\begin{eqnarray}
    \Gamma_{\rm DT} = \frac{3m}{2} \frac{\left[ \ln{\left(1+\dfrac{t_i}{t_d} \right)}\right] ^m}{\left[ \ln{\left(1+\dfrac{t_i}{t_d}\right)} + \frac{3}{2} \ln{\left(\dfrac{1+z_i}{1+z}\right)}\right]^{m+1}} H(z)\, \rho_B(z),
    \label{eq:gamma_DT}
\end{eqnarray}
where $m=2 \, (n_B+3)/(n_B+5)$, and $n_B$ is the spectral index corresponding to the primordial magnetic field.  The physical decay time scale for turbulence is $t_d$ and the time at which decaying turbulence becomes dominant is $t_i$, and these are related as $t_i/t_d \simeq 14.8(B_0/{\rm nG})^{-1}(k_D/{\rm Mpc}^{-1})^{-1}$, where $B_0$ is the present day magnetic field and $k_D$ is the damping scale \citep{Chluba15}.

\subsection{Cooling due to dark-matter baryon interaction}
\label{th:cold_IGM}
As mentioned in the previous section, the excess radio background changes $T_{\gamma}$, which in turn affects the 21-cm signal. Similarly, it has been shown in a few recent studies that the interactions between the cold dark matter particles and baryons could help the IGM to cool faster than the standard adiabatic cooling \citep{Barkana18Nature,PhysRevLett.121.011102,Munoz18,Liu2019prd}. The interaction could be Rutherford like where interaction cross-section depends on the relative velocity, $v$, as,  $\sigma = \sigma_0 (v/c)^{-4}$ \citep{Barkana18Nature}. The milli-charged dark matter model is one such kind of interaction model \citep{Munoz18}. 
In case of Rutherford like interaction, the energy transfer rate to baryons from dark-matter due to such interaction can be written as \citet{Munoz15}, 
\begin{eqnarray}
    \frac{dQ_b}{dt} = \frac{2 m_b \rho_{\chi} \sigma_0 e^{-r^{2} / 2} (T_{\chi} - T_{\rm K}) k_B c^4}{(m_b + m_{\chi})^2 \sqrt{2\pi} u^{3}_{th}} \nonumber \\
    + \frac{\rho_{\chi}}{\rho_m} \frac{m_{\chi} m_b}{m_{\chi} + m_b} V_{\chi b} \frac{D(V_{\chi b})}{c^2}.
    \label{eq:dQ_b}
\end{eqnarray}
Here, $m_\chi$, $m_b$ and $\rho_\chi$, $\rho_b$ are the masses and energy densities of dark matter and baryon, respectively, and $u_{th}^2 = k_B(T_b/m_b + T_\chi/m_\chi$) is the variance of the thermal relative velocity of dark-matter and baryons. The relative velocity between dark matter and baryon, $(V_{\chi b})$ also evolves with redshift and can be given as,
\begin{equation}
\frac{dV_{\chi b}}{dz} = \frac{V_{\chi b}}{1+z} + \frac{D(V_{\chi b})}{H(z)(1+z)}
\label{eq:vxb}
\end{equation}
with,
\begin{equation}
D(V_{\chi b})  = \frac{\rho_m \sigma_0 c^4}{m_b + m_\chi} \frac{1}{V^2_{\chi b}} F(V_{\chi b}/u_{th}).
\end{equation}
The function $F$ is given by,
\begin{equation}
F(r) = erf \Big( \frac{r}{\sqrt{2}} \Big) - \sqrt{ \frac{2}{\pi}} r e^{-r^2/2},
\end{equation}
with $F(0) = 0$ and $F(\infty)= 1$. It can be seen from Eq.~\ref{eq:dQ_b} that the first term depends on the temperature difference between two fluids. In case of the cold dark matter scenario, the temperature of dark-matter is likely to be lower than the IGM, and hence this term helps to cool the baryon fluids. The second term arises due to the friction between dark matter and baryon fluids as they have different velocities, and hence, both the fluids get heated up irrespective of their own temperature and depending on their relative velocity, $V_{\chi b}$.
However, this term is subdominant during the cosmic dawn, and as a result of which the IGM temperature decreases due to this Rutherford like interaction.

\section{Global signal}
\label{sec:global}
We have already described current theoretical understanding of cosmic dawn along with the various physical processes which play important roles in shaping the 21-cm signal.
Further, different processes mentioned in the above sections dominate at separate regimes in the evolution of 21-cm signal. For example, once the star formation begins, the Ly-$\alpha$ photon flux generated from the first generation of stars starts to couple the spin temperature to the gas kinetic temperature. Since, the gas temperature is below the CMBR temperature, one expect the 21-cm signal in absorption. The shape and strength of the absorption depends on the Ly-$\alpha$ coupling strength. As soon as the Ly-$\alpha$ coupling saturates, the shape of the 21-cm signal is governed by the other heating processes. Note that, the impact of these processes on the global signal is not sharply distinguished, hence there could be a overlap, and depends on the various efficiency parameters associated with each processes. 

\begin{figure*}
    \centering
    \includegraphics[width=\linewidth]{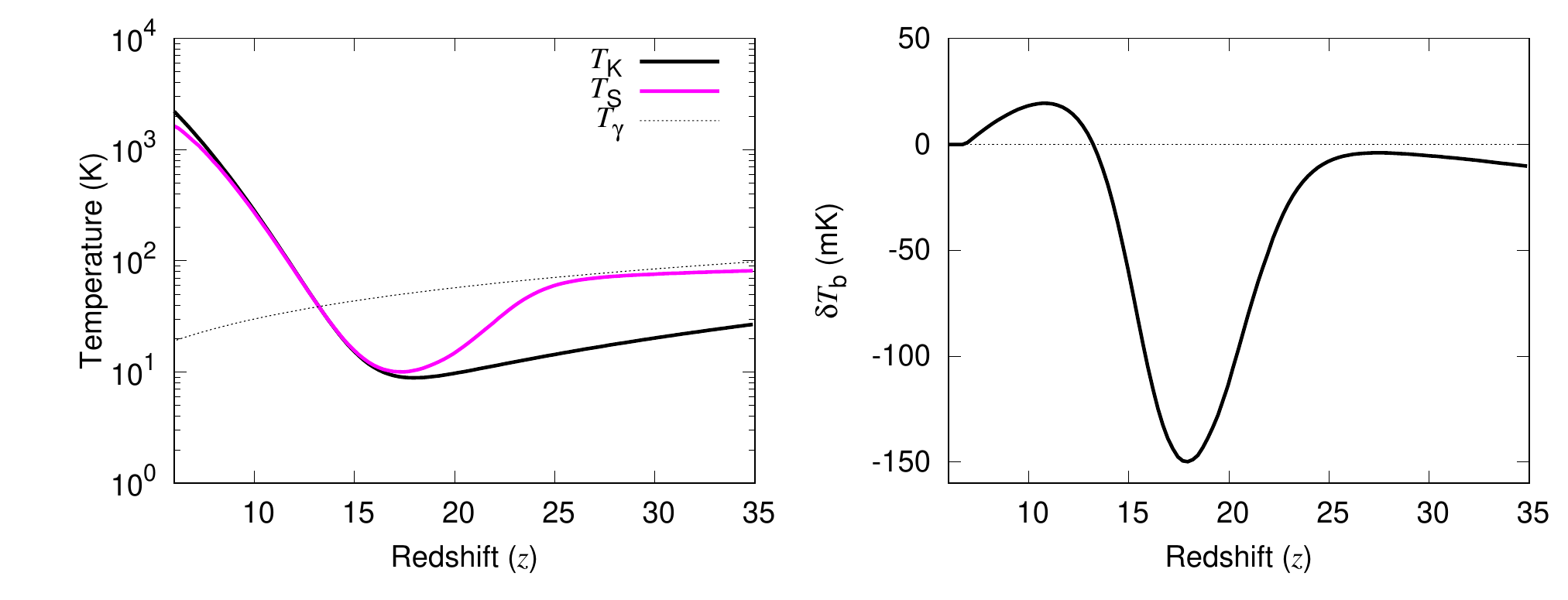}
    \caption{Left panel: The evolution of $T_{\gamma}, T_s, T_{\rm K}$ with redshift is shown. In this scenario, Ly-$\alpha$ coupling and X-ray heating are considered. The source model considered here has a parameter set of $f_*=0.1, f_X=1, T_{\rm vir}=10^4$~K. Right panel: The corresponding global 21-cm signal is shown for the same set of parameters.}
    \label{fig:tgt21xray}
\end{figure*}

One such scenario has been shown in Fig.~\ref{fig:tgt21xray}, where the coupling due to Ly-$\alpha$ photons and heating of IGM due to X-ray photons are considered \citep[this plot is generated using the analytical code used in][]{Nebrin19}. For this particular result, only the atomic cooling halos with virial temperature, $T_{\rm vir}=10^4$~K are considered for star formation with the star formation efficiency of $f_*=0.1$, where $f_*$ is defined as the fraction of baryons residing within the stars in a galaxy. Further, the X-ray efficiency is considered to be $f_X=1$, which means that sources produce a total of $f_X \times 3.156 \times 10^{48} \rm erg$ energy per stellar mass in the X-ray band which we assume to span $0.1-10$ keV. (see Eq.~\ref{eq:Lx}). The left panel shows the variation in $T_{\rm K}$, $T_s$, and $T_{\gamma}$ by black solid, magenta solid, and black dotted curves respectively, whereas in the right panel, the corresponding brightness temperature is shown by black solid curve. 
It is clear from Fig.~\ref{fig:tgt21xray}, $T_s$ starts to decouple from $T_{\gamma}$ at $z\sim 25$, and finally couples with $T_{\rm K}$ by $z\sim 17$ due to the presence of Ly-$\alpha$ coupling. This leads $T_s$ to reach a minimum of $\sim 10$~K, which produces an absorption depth of $\sim 150$~mK (shown in the right panel). Then due to the X-ray heating, $T_{\rm K}$ as well as $T_s$ cross the CMBR temperature at $z\sim13$. Hence, the presence of Ly-$\alpha$ coupling along with X-ray heating can successfully produce 21-cm absorption profile of depth $100-200$~mK which is generally required in a standard cosmological model.

\begin{figure}[ht]
    \centering
    \includegraphics[width=\columnwidth]{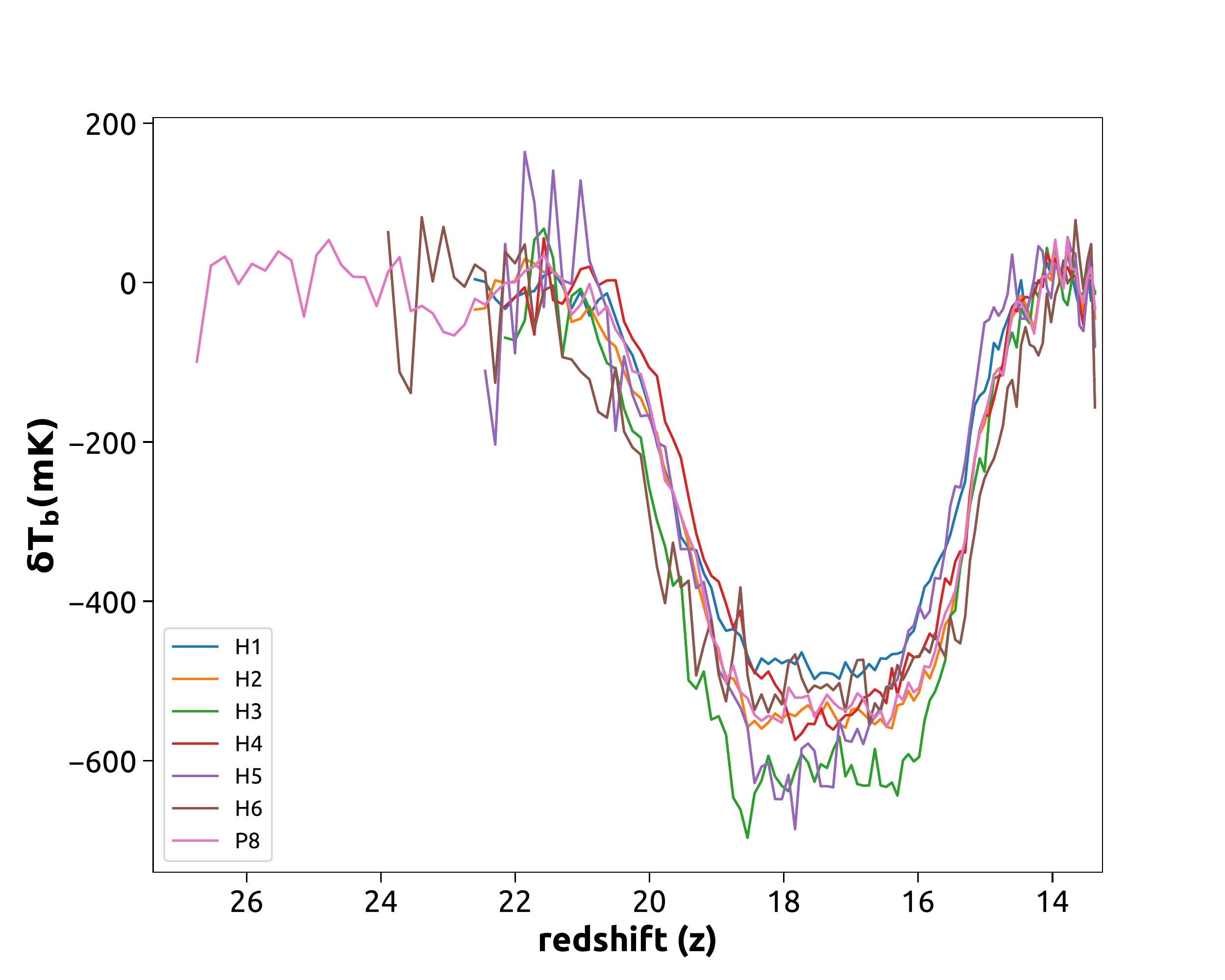}
    \caption{EDGES observation of the global 21~cm signal. The different hardware configuration that were used for the observations are denoted by H1, H2, H3, ....P8. Note the $-500$ mK absorption trough (around $z \sim 17$) which is twice that of the prediction coming from standard galaxy formation model.  The data, used here for plotting, is publicly available at \href{http://loco.lab.asu.edu/edges/edges-data-release/}{http://loco.lab.asu.edu/edges/edges-data-release/}. }
    \label{fig:EDGES}
\end{figure}
However, there is a possible detection of the global HI 21-cm absorption signal by the Experiment to Detect the Epoch of Reionization Signature (EDGES) 
which shows that the detected signal has an absorption depth of $0.5^{+0.5}_{-0.2}$~K centred at frequency $78\pm1$~MHz or redshift $z \sim 17$ as can be seen from Fig.~\ref{fig:EDGES} \citep{EDGES2018}. Note that, there are concerns regarding the foreground removal and unaccounted systematics that could lead to the misinterpretation of the signal \citep[see][for deailed description]{Hills_2018,Bradley_2019, singh2019, Sims2020}. 
A lot of efforts which are fundamentally different have been put up in order to understand this globally averaged signal detected by EDGES \citep{Barkana18Nature,PhysRevLett.121.011102,Munoz18,Liu2019prd,Feng_2018,Ewall_2018,Fialkov_2019}. One such possible explanation is the interaction between cold dark-matter and baryon that we have discussed in Sec.~\ref{th:cold_IGM}. 
\begin{figure}[ht]
    \centering
    \includegraphics[width=\columnwidth]{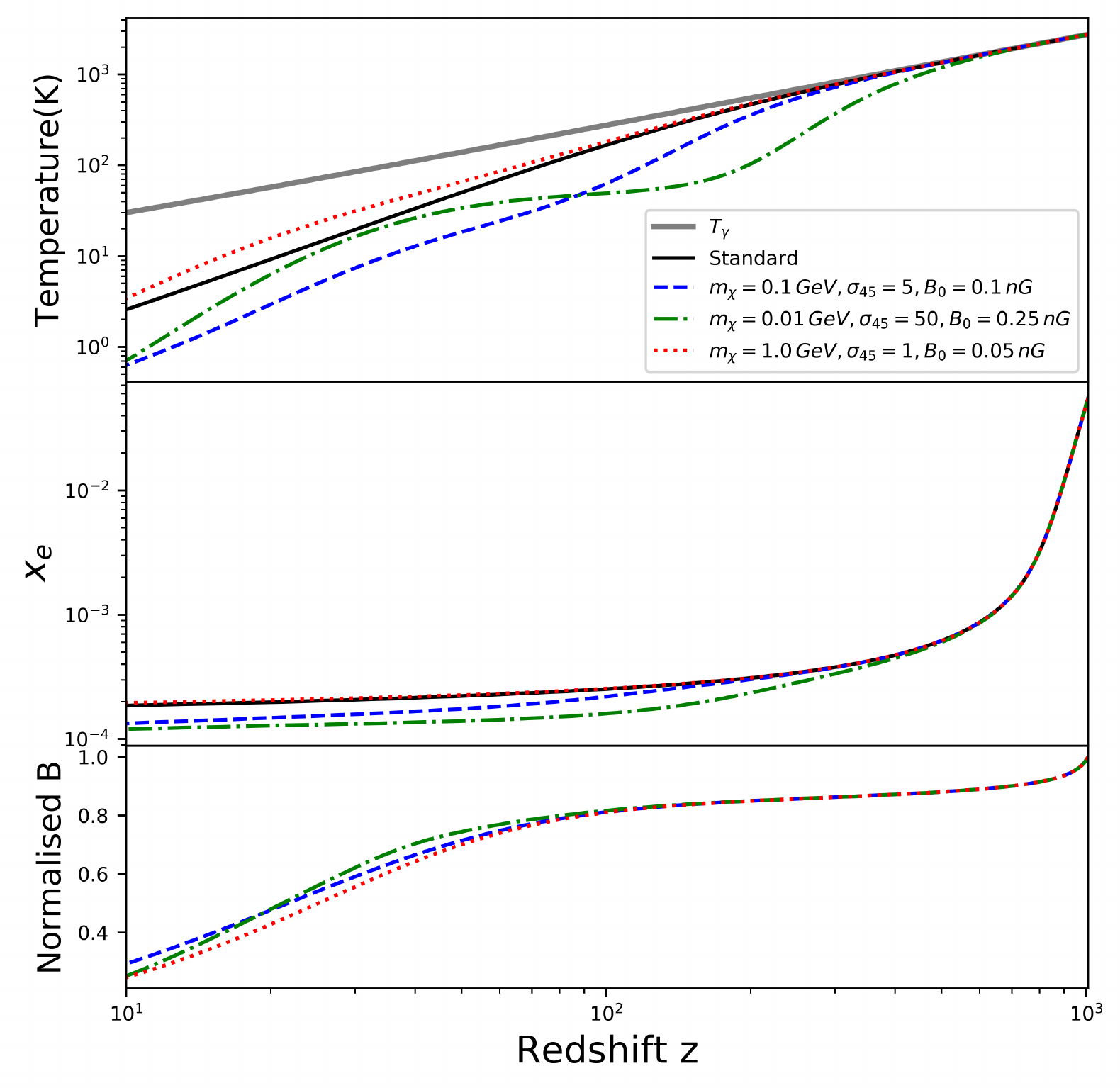}
    \caption{The upper and middle panels show the IGM kinetic temperature $T_{\rm K}$ and residual free electron fraction $x_e$ as a function of redshift in presence of the primordial magnetic field and dark-matter baryon interaction. The solid grey line in the top panel shows the CMBR temperature, $T_{\gamma}$. The lower panel shows the normalised magnetic field $B(z)$, and we refer the reader to \citet{Bera_2020} for more details. The figure is adopted with permission from \citet{Bera_2020}.}
    \label{fig:DM_PMF}
\end{figure}
Note that, such kind of interaction cools the IGM, hence it decides the depth of the absorption signal. However, the width of the absorption signal depends on the heating mechanisms those are effective during cosmic dawn. One such scenario of dark-matter baryon interaction along with the magnetic field heating is considered in \citet{Bera_2020}, and their results are shown in Fig.~\ref{fig:DM_PMF}. 
This plot shows the variation in gas temperature and the impact on the ionization fraction for different parameters shown in the legend. 

The upper panel of Fig.~\ref{fig:DM_PMF} shows that the primordial magnetic field and DM-baryonic interaction together introduces a `plateau like feature' in the redshift evolution of the IGM temperature. The plateau like feature is more prominent for a set of parameters $m_{\chi}=0.01\, {\rm GeV}$,  $\sigma_{45}=50$  and $B_0=0.25 \, {\rm nG}$.  In this case, the cooling rate due to the DM-baryonic interaction and heating rate due to the primordial magnetic field compensates each other for a certain redshift range, but at lower redshifts the heating due to the primordial magnetic field becomes ineffective due to the decaying of magnetic field.   
The middle of Fig.~\ref{fig:DM_PMF}  shows the residual electron fraction, $x_e$ as a function of redshift respectively. As discussed in the previous cases, the residual electron fraction $x_e$ gets suppressed  when both the DM-baryonic interactions and primordial magnetic field are active.  The EDGES absorption spectra show that the IGM temperature is rising  at  redshifts $z \lesssim 17$. 
However, from the figure, it is clear that the magnetic field can heat the IGM significantly but heating is not much efficient at low redshift, hence can not explain the heating part of the EDGES absorption profile \citep[for details see][]{Bera_2020}. Note that, the magnitude of magnetic field, and the dark-matter baryon interaction parameters can be constrained using the global 21-cm signal as discussed in \citet{Minoda19} and \citet{Bera_2020}. Further in the lower panel of Fig.~\ref{fig:DM_PMF}, normalised magnetic field, $B(z)$ is plotted for the same set of parameters. The details can be found in \citet{Bera_2020}.

\begin{figure*}
    \centerline{\includegraphics[width=\columnwidth]{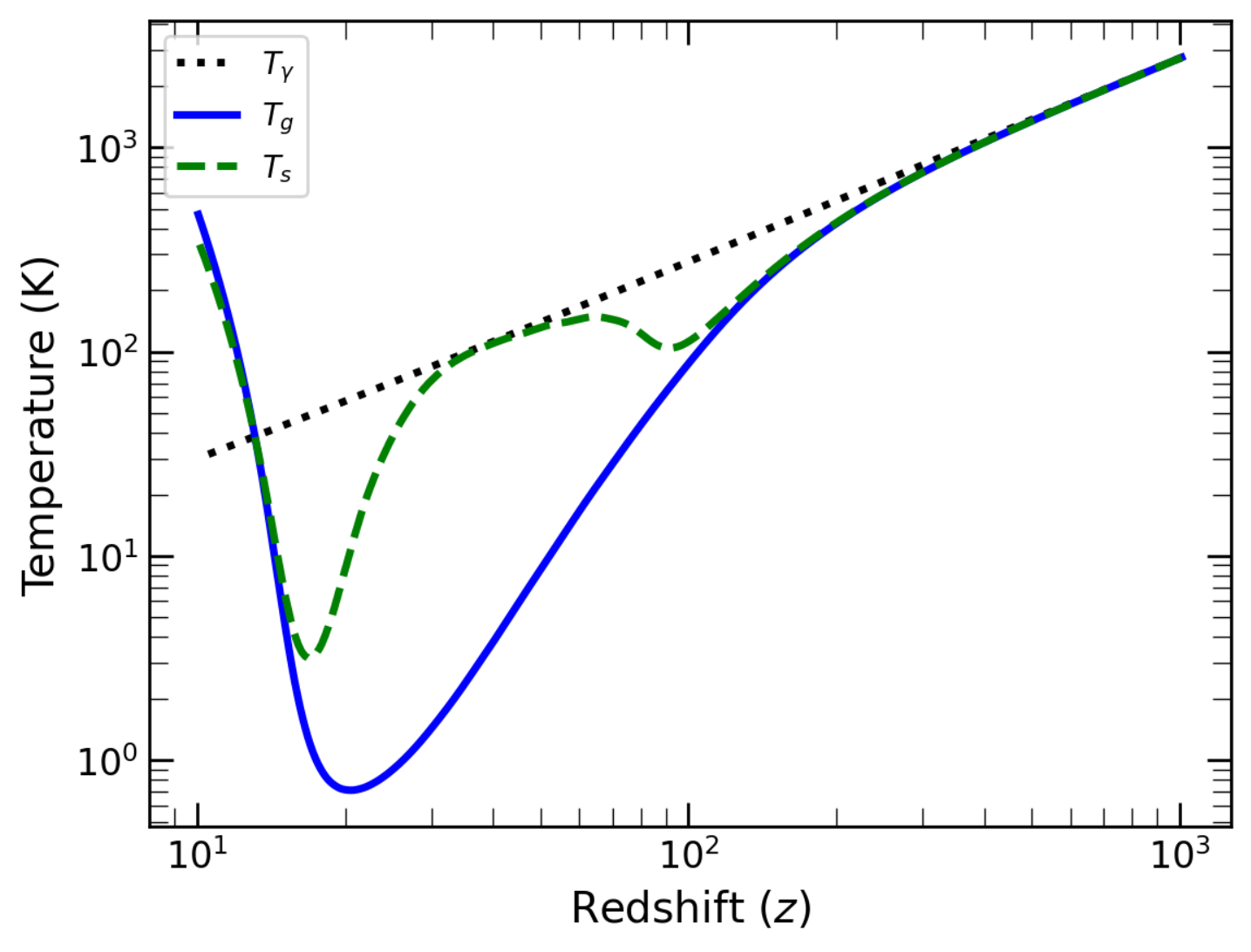}\includegraphics[width=\columnwidth]{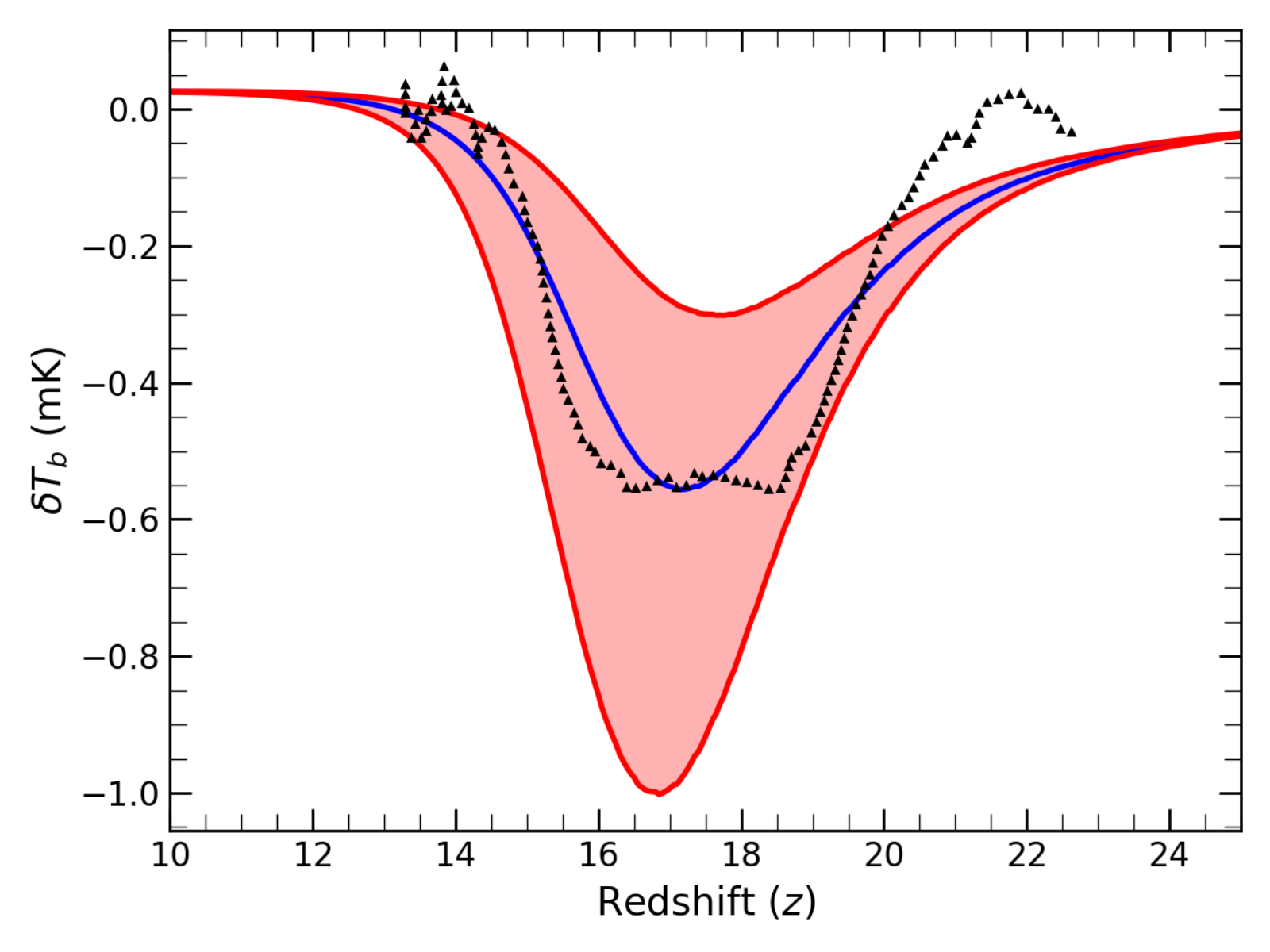}}
    \caption{Left panel: The CMBR temperature $T_{\gamma}$(black dotted), gas temperature $T_{\rm K}$(blue solid curve), and spin temperature $T_s$(green dashed curve) are plotted in presence of dark-matter baryon interaction of $(m_{\chi}/{\rm GeV}, \sigma_{45})=(0.1, 2)$. Right panel: The corresponding global 21-cm absorption profile is shown by blue solid curve. By varying the model parameters, other 21-cm profiles can be obtained those are within the upper and lower bounds of the absorption depth reported by EDGES, as shown by the red shaded region. The best fit 21-cm profile of EDGES is also plotted by black triangles for reference.}
    \label{fig:Ts_Tg_T21_CR}
\end{figure*}

As already mentioned, another possible source of heating could be cosmic rays which was considered in \citet{Bera2022}.
The result of such model is shown in Fig.~\ref{fig:Ts_Tg_T21_CR} where, along with the Ly-$\alpha$ coupling, only the heating by cosmic ray particles are considered in presence of dark-matter baryon interaction, and no X-ray heating is considered. 
In the L.H.S plot of Fig.~\ref{fig:Ts_Tg_T21_CR}, CMBR temperature, gas temperature and spin temperature are plotted by black dotted, blue solid and green dashed curves respectively for a parameter set of ($q$, $\epsilon_{\rm III}$, $\epsilon_{\rm II}$, $f_{\alpha, \rm III}$, $f_{\alpha, \rm II}$, $m_{\chi}/{\rm GeV}$, $\sigma_{45}$) = ($2.2, 0.05, 0.16, 0.1, 1, 0.1, 2$). Here, $q$, $\epsilon_{\rm III}$, and $\epsilon_{\rm II}$ denote spectral index of cosmic ray spectra, efficiency of cosmic rays generated in Pop~III and Pop~II stars respectively. Further, $f_{\alpha}$ is the normalization parameter which takes care of the uncertainty in the escape of the Ly-$\alpha$ photons and properties of first stars. The other two parameters, $m_{\chi}$, and $\sigma_{45}$ represent the dark-matter mass and interaction cross-section, $\sigma_{45}=\sigma_0/(10^{-45} \rm m^{-2})$. These parameters are explored more elaborately in \citet{Bera2022}. 
It can be seen from Fig.~\ref{fig:Ts_Tg_T21_CR}, that the gas kinetic temperature $T_{\rm K}$ and CMBR temperature $T_{\gamma}$ are coupled to each other upto redshift $z \gtrsim 200$ through Compton scattering process. Afterwards, the presence of dark-matter baryon interaction along with adiabatic cooling results in sharp fall of $T_{\rm K}$. Due to collisional coupling, $T_s$ follows $T_{\rm K}$ upto $z\sim 100$. As soon as the collisional coupling becomes weak due to the lower IGM temperature and density, $T_s$ starts to follow $T_{\gamma}$. As discussed before, $T_s$ again decoupled from $T_{\gamma}$ due to the presence of Ly-$\alpha$ photons around $z \sim 30$.
As the contribution of Ly-$\alpha$ photons from Pop~III stars are also considered in this model, it helps to initiate the coupling of $T_s$ to $T_{\rm K}$ at an earlier redshift. Afterwards, $T_{\rm K}$ as well as $T_S$ reach a minimum  at $z \sim 20$. The cosmic rays protons generated from the first generation of stars are enough to heat the ambient gas, and finally $T_s$ and $T_{\rm K}$ cross the CMBR temperature at redshift $z \sim 12$. 
In the R.H.S plot, the corresponding 21-cm absorption signal is plotted by the blue solid curve. The red shaded region is achieved by varying the efficiencies of cosmic rays generated from Pop~III and Pop~II stars. It is clear that the entire range of the absorption depth reported by EDGES can be described by the Ly-$\alpha$ coupling along with cosmic rays heating. The best-fit profile of EDGES is also plotted by black triangles for reference. The heating due to cosmic rays help the absorption signal to rise sharply which can be seen Fig.~\ref{fig:Ts_Tg_T21_CR}.
It is clear from this figure that cosmic ray is a potential source of heating for a steeper absorption. 

Note that, Fig.~\ref{fig:Ts_Tg_T21_CR} is shown for a set of parameters that quantifies the amount of Ly-$\alpha$ flux input and heating due to cosmic rays. Hence, the evolution of 21-cm absorption signal could vary depending on the parameters related to the various physical processes considered in different models.
Further, the detection of 21-cm signal may provide constraint on the early sources and their subsequent feedbacks.

\begin{figure*}
    \centering
    \includegraphics[width=\linewidth]{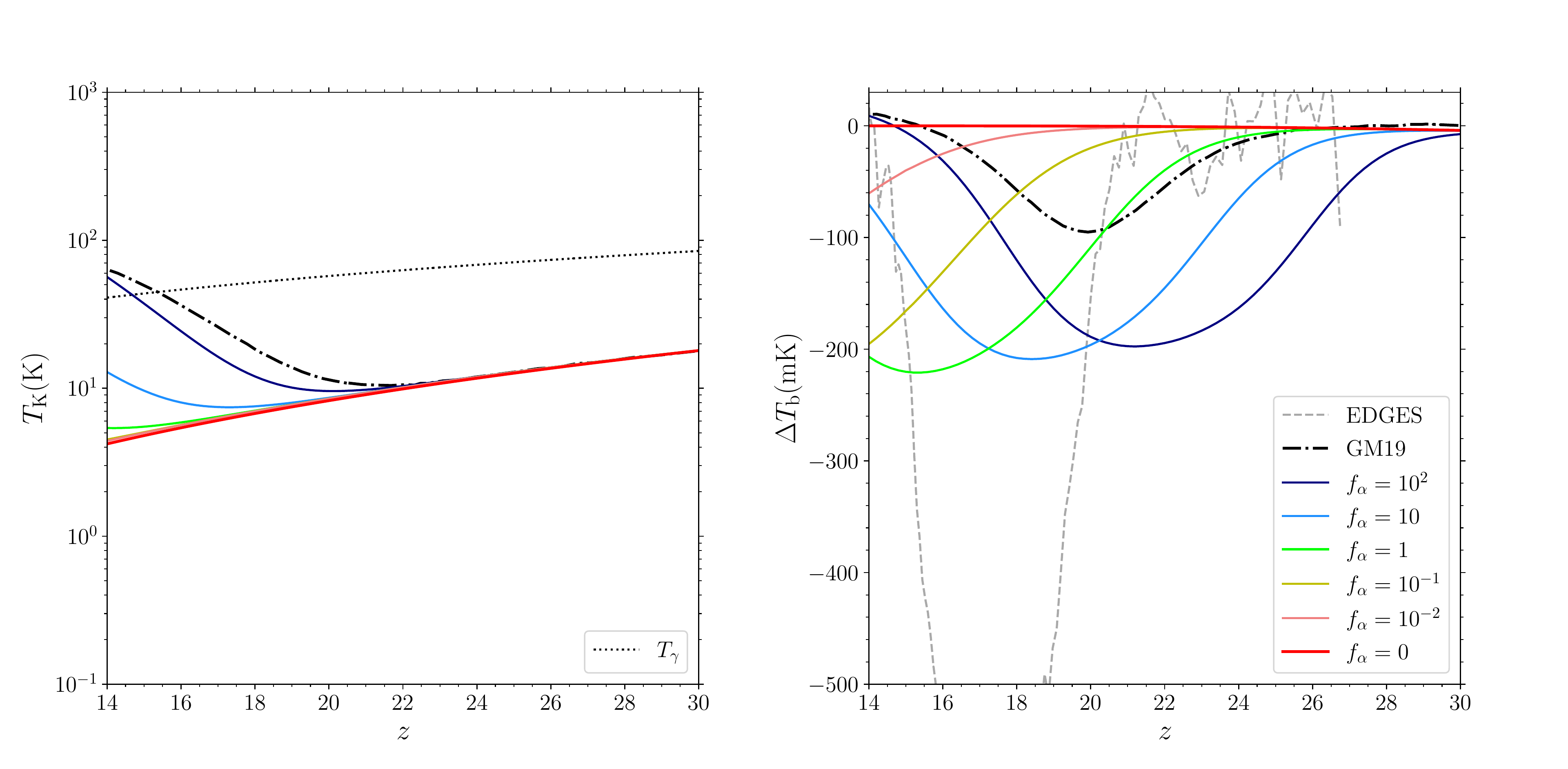}
    \caption{Left panel: The evolution of $T_{\rm K}$ with redshift are shown for different $f_{\alpha}$ shown in the legend. Here, $f_{\alpha}$ is the scaling factor that controls the coupling and heating due to Ly-$\alpha$ photons, and the detail can be found in \citet{mittal21}. Right panel: The corresponding differential brightness temperatures are shown. The grey dashed line denotes the EDGES detection. The figure is adopted with permission from \citet{mittal21}}
    \label{fig:t21_lyheating}
\end{figure*}
People have also studied another heating process such as heating due to Ly-$\alpha$ photons \citep{Chen_2004,Furlanetto06MNRAS,ghara2020,mittal21}. One such mechanism during cosmic dawn is shown in Fig.~\ref{fig:t21_lyheating}, where they used a parameter $f_{\alpha}$ which varies the strength of the Ly-$\alpha$ radiation \citep{mittal21}. As a result, the heating of the IGM due to Ly-$\alpha$ photons gets affected significantly (shown in the left panel), and in turn the depth and duration of differential brightness temperature changes (shown in the right panel). It is clear form Fig.~\ref{fig:t21_lyheating} that a strong Ly-$\alpha$ background is required, though not sufficient, in order to fit with the EDGES observed signal. 
One can explore the efficiency of heating due to different processes such as CRs, X-rays, Ly-$\alpha$, or any other heating processes. 

It is worth mentioning that a recent observation by Shaped Antenna measurement of the background RAdio Spectrum (SARAS 3) \citep{Saurabh_2021} claims a null detection of the global 21-cm absorption signal and they rule out the EDGES detection with 93.5 \% confidence . As they used a completely different experimental setup and highlighted that the deeper absorption profile of EDGES may not be of astrophysical origin, one needs to wait for other experiments to settle this issue. However, other aspect of 21-cm signal namely the power spectrum can be a complementary probe of the HI distribution during cosmic dawn that SKA is aiming to probe. We discuss this aspect in the following section.


\section{Power spectrum} 
\label{sec:ps}
It is expected that the IGM kinetic temperature takes over the background CMBR temperature (i.e., $T_S>>T_\gamma$) during the EoR and, thus, the spatial fluctuations in the signal during EoR are mainly determined by the fluctuations in the neutral hydrogen density. On the contrary, the ionization remains low during cosmic dawn and fluctuations in the spin temperature distribution is expected to dominate the HI power spectrum during cosmic dawn given that the radio background remains uniform. It should be realized that mechanisms that have significant impacts on the amplitude and spatial structures of $T_\gamma$ and $T_S$ are important for the study of the 21-cm signal power spectrum during cosmic dawn. In section \ref{sec:rad_par} \& \ref{sec:heat_source}, we discussed processes that  have significant impact on the spin temperature and HI 21-cm signal from cosmic dawn.

\begin{figure}
 \centering
\includegraphics[scale=0.6]{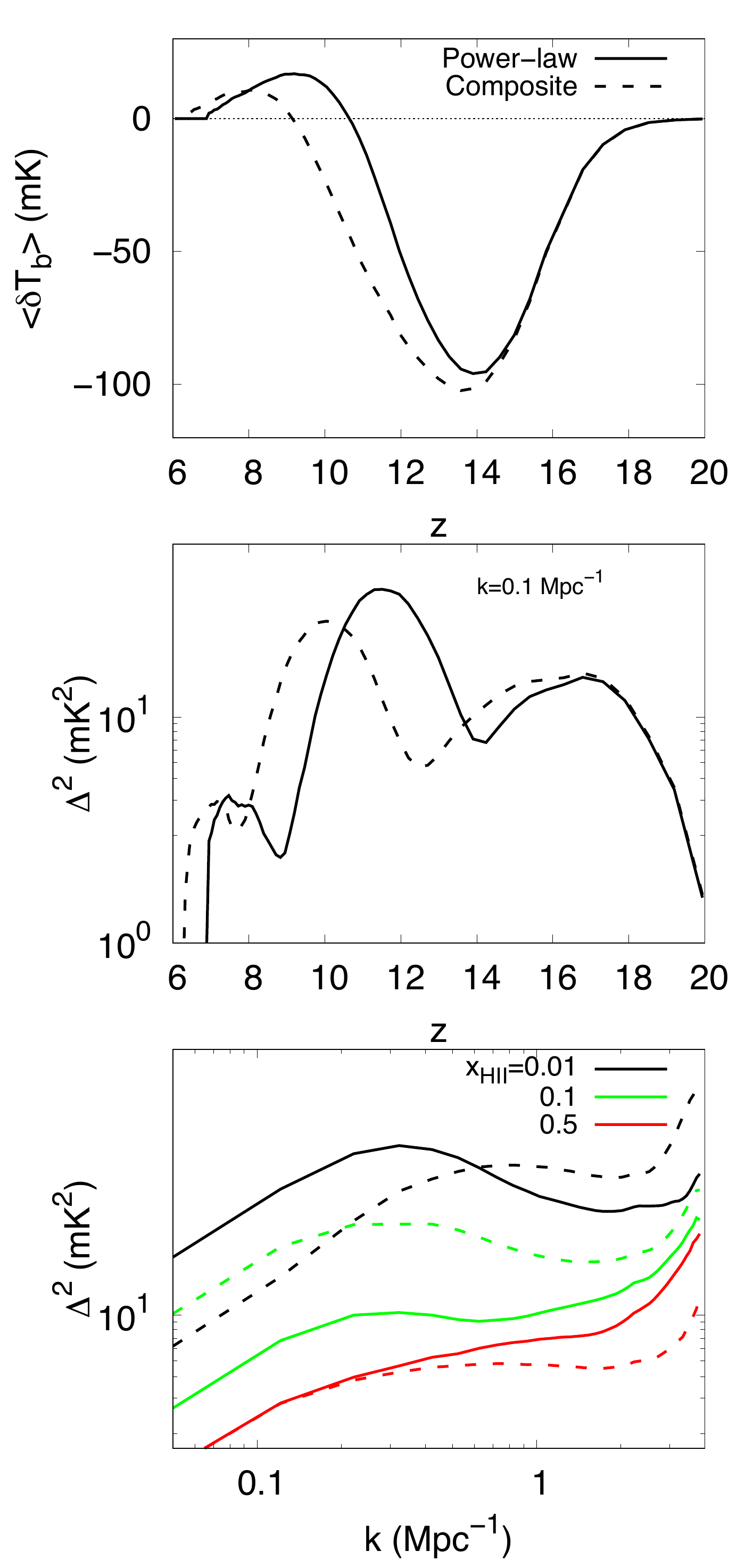}
    \caption{Top Panel: Redshift evolution of the volume averaged brightness temperature of the 21-cm signal; for two different source models, dashed curve presents the composite spectrum derived from the observation with MAXI telescope while the solid curve corresponds to power-law spectrum with a spectral index of 1.5. Middle Panel: Redshift evolution of the large-scale power spectrum of the 21-cm signal; The curves correspond to scale $0.1 ~\rm Mpc^{-1}$ . Bottom panel: The power spectrum of 21-cm signal as a function of scale at different stages of reionization for the two different source models; dashed curves present the composite spectrum, while the thick curves correspond to the power-law spectrum. The figure is adopted with permission from \citet{2019MNRAS.487.2785I}.}
   \label{fig:image_pszk}
\end{figure}

Theoretical studies such as \citet[][]{2019MNRAS.487.2785I, 2019MNRAS.487.1101R, ghara15c} show that the measured visibilities and 21-cm power spectrum are very sensitive to the radiation backgrounds that majorly depend on the underlying astrophysical sources. For example, at the same global averaged ionization states of the Universe, the power spectrum of the 21-cm signal brightness temperature would be distinctly different between two scenarios. Such scenarios are shown in the middle and bottom panels of Fig.~\ref{fig:image_pszk} where one is driven by high-mass X-ray binaries (dashed curves) and the second one is driven by a power-law X-ray sources such as mini-quasars (solid curves). In particular, the large-scale power spectrum which has a higher detectability in a radio interferometric observation is more sensitive to the radiation backgrounds such as UV, X-rays, Ly-$\alpha$. For example, the redshift evolution of the large-scale power spectrum is expected to show a `three-peak' feature where the peaks from high redshift to low redshift are due to fluctuations in Ly-$\alpha$ coupling, X-ray heating and ionization (see Fig.~\ref{fig:image_pszk}) \citep{baek10, ghara15a}. This sensitive nature of the power spectrum due to change in source model provides the opportunity to tightly constrain the properties of these early sources (such as high-mass X-ray binaries, mini-quasars) using the interferometric observations of cosmic dawn 21-cm signal. 

Besides heating, processes that might cool the IGM gas other than the cosmological expansion of the Universe also play important roles in shaping the cosmic dawn 21-cm power spectrum \citep[][]{2018PhRvL.121a1101F, 2018Natur.557..684M, PhysRevLett.121.011102}, like in case of global 21-cm signal discussed in Sec.~\ref{sec:global}.
In addition, the presence of an excess radio background to the CMB in the form of both the uniform and fluctuating can also significantly alter the cosmic dawn 21-cm power spectrum \citep[][]{2020MNRAS.499.5993R, 2020MNRAS.498.4178M, 2021arXiv210813593G}. Further, the HI 21-cm signal also depends on the nature of dark matter particles \citep[][]{2014MNRAS.438.2664S, Nebrin19}.  In principle, one should consider all possible mechanisms that have non-negligible contributions while modelling the cosmic dawn 21-cm signal. However, it is not straightforward to separate out the relative contribution of a process using 21-cm measurements unless its impact on the signal is unique.

In addition to the $T_\gamma$ and $T_S$ dependencies, the cosmic dawn power spectrum also crucially depends on the different line of sight effects such as the redshift space distortion (RSD), light-cone effect. Studies such as \citet{Bharadwaj_2004, mao12, Jensen13, majumdar13} show that RSD can boost the power spectrum of cosmic dawn by a factor of $\approx 2$ when fluctuations can be assumed linear. The spin temperature fluctuations during the cosmic dawn suppresses the impact of RSD on the power spectrum \citep[see e.g.,][]{ghara15a, 2021MNRAS.506.3717R}. Further, it has been shown that  the light-cone effect can have significant impact on the large HI power power spectrum during initial and late stages of reionization \citep[see e.g.,][]{Datta2012b, datta14}, while the effect can enhance (suppress) the large-scale power spectrum by up to a factor of $\sim 3$ (0.6) at different stages of cosmic dawn for a non-uniform spin temperature \citep[see e.g.,][]{ghara15b}. 

\subsection{Current upper limits on HI power spectrum}
The ongoing radio interferometric experiments that aim to detect this faint signal from the cosmic dawn and EoR have, to date, only provided upper limits on the power spectrum. The first upper limit on the  dimensionless power spectrum of the signal at  redshift $8.6$  was obtained by GMRT which reach a 2$\sigma$ value of $\Delta^2(k) < (248)^2~ \rm mK^2$ for a $k$-scale $0.5 ~h ~\rm Mpc^{-1}$ \citep{paciga13}. A recent study using the GMRT obtains a $2\sigma$ upper limit of $\sim (73)^2 \, {\rm K}^2$  on $\Delta^2(k)$ at a much smaller scale  at $k=1.59 \, {\rm Mpc}^{-1}$ \citep{Pal21}. The best upper limit on the power spectrum for $k$-scale $0.37 ~h ~\rm Mpc^{-1}$ at redshift $8.37$ as obtained by PAPER interferometer is $(200 \rm mK)^2$ \citep{2019ApJ...883..133K}. On the other hand, the best upper limit from LOFAR interferometer is at $k$-scale  $0.075 ~h ~\rm Mpc^{-1}$ with 2$\sigma$ value of  $(73)^2 \rm ~mK^2$ at redshifts $9.1$ \citep{martens20}.  The best upper limit obtained by MWA interferometer is at redshift $6.5$ with a 2$\sigma$ value $\Delta^2(k = 0.14 ~h ~\rm Mpc^{-1})\approx (43)^2 \rm ~mK^2$. Recently, HERA interferometer published 2$\sigma$ upper limits of  $ {(30.76)}^{2}\ {\mathrm{mK}}^{2}$ at $k = 0.192$ $ ~h ~\rm Mpc^{-1}$ at $z = 7.9$ \citep{2022ApJ...925..221A} which is the best upper limit so far. Note that, these are not the complete set of upper limits. These experiments also produced  upper limits at other redshifts which includes redshift as large as $z\sim 20-25$ and post reionization epochs \citep{2019ApJ...884....1B, 2017ApJ...838...65P, 2019MNRAS.488.4271G,2020MNRAS.499.4158G, 2019AJ....158...84E, Arnab_GSB_2019a}.

\begin{figure*}
\centering
\includegraphics[width=\textwidth]{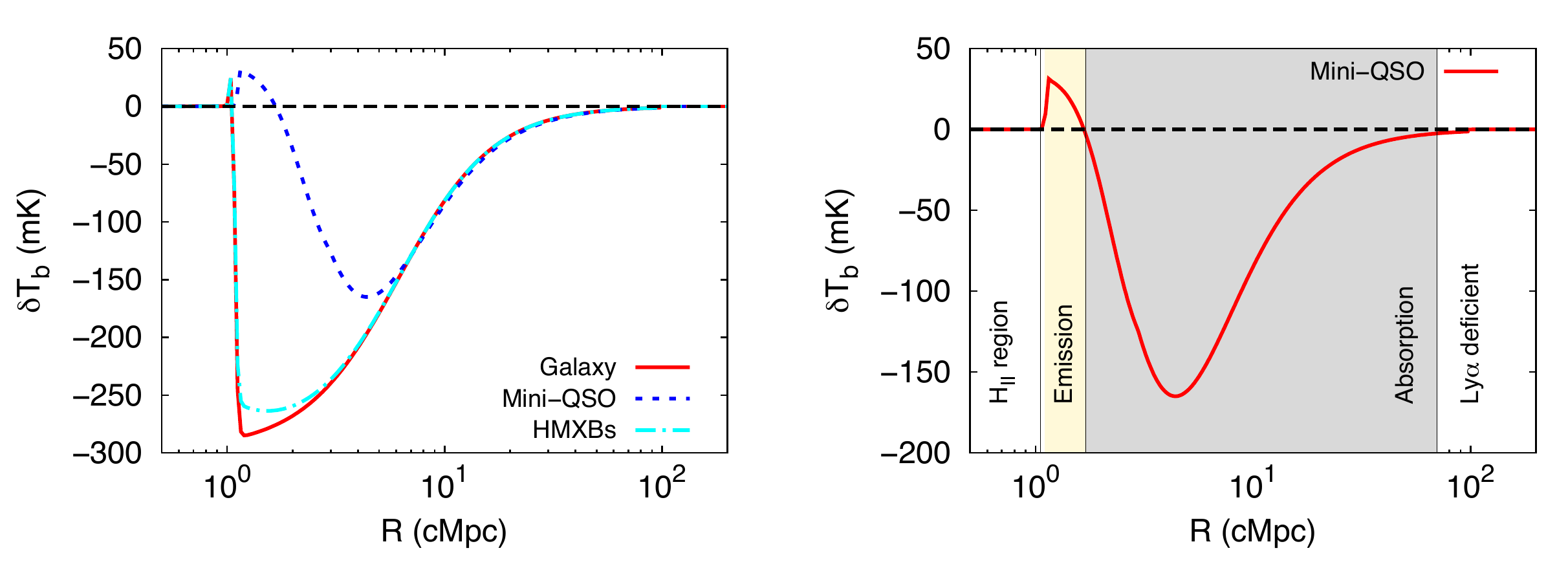}
    \caption{Left Panel: The radial brightness temperature pattern as a function of the radial distance R from the centre of the model sources. Different lines represent different types of source models where we keep the stellar mass of the source fixed to $10^7 ~\MSUN$. Right panel: Four separate regions around a mini-QSO type source. The plot is adopted with permission from \citet{ghara16}.}
   \label{image_tbprof}
\end{figure*}

\begin{figure}[ht]
\centering
\includegraphics[scale=0.5]{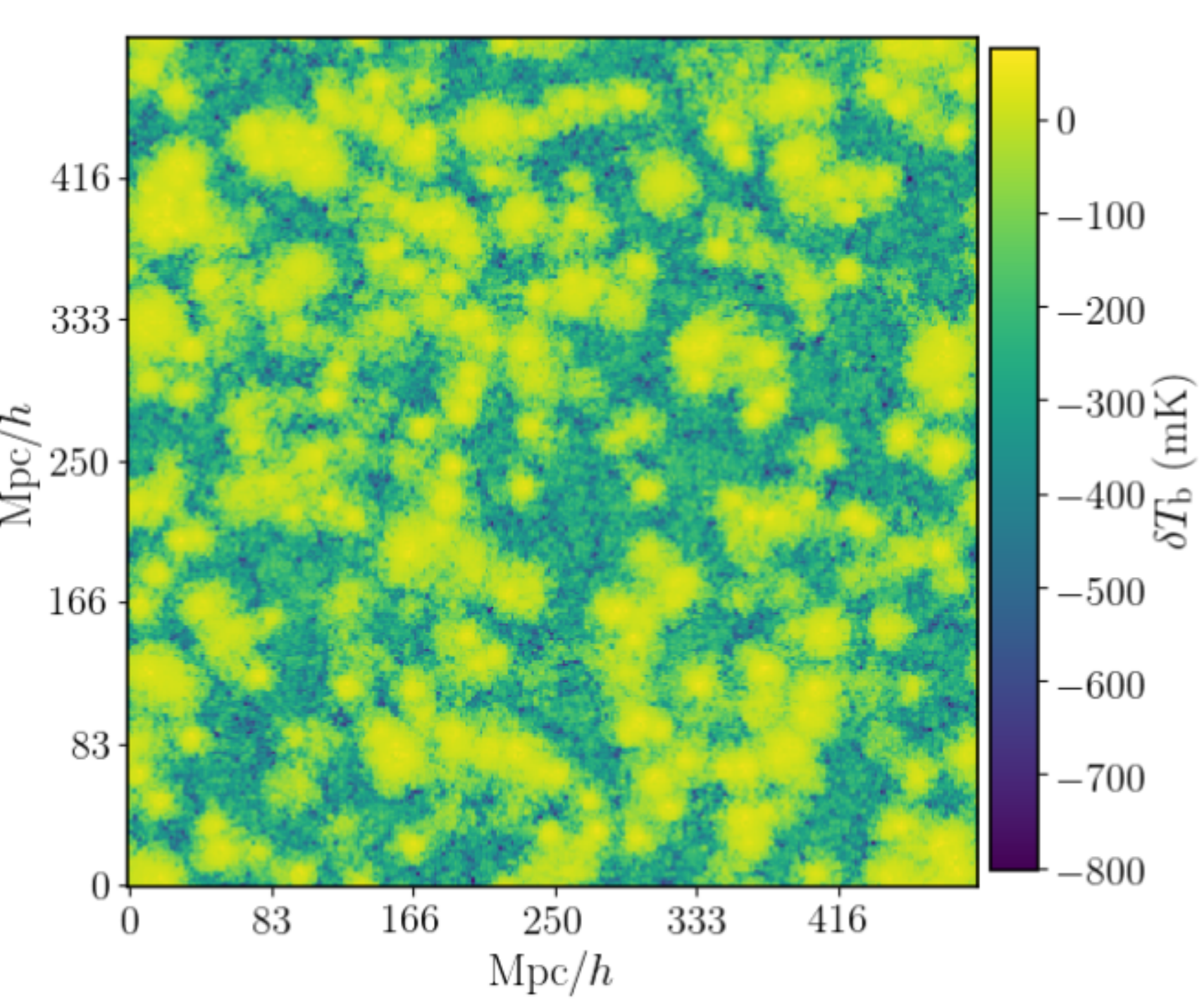}
    \caption{A brightness temperature slice at redshift $z\approx 9.1$ with ionization fraction 0.01. The 21-cm signal is dominated by spin temperature fluctuations. For more details of the source model used to generate this map, we refer the reader to \citet{2020MNRAS.493.4728G}, and this figure is adopted with their permission. }
   \label{image_tbmap}
\end{figure}

\section{Parameter estimation}
\label{param_est}
Accurately measured HI 21-cm signal can put constraint on various astrophysical parameters related to the physics that determines the 21-cm signal along with the cosmological parameters.
Recently a huge body of work \citep{ 2016MNRAS.459.2342M, 2017MNRAS.465.4838G, 2017MNRAS.468..122H, 2017ApJ...848...23K, 2018MNRAS.477.3217G,2018MNRAS.475.1213S, 2019MNRAS.484..933P, 2021MNRAS.503.4551G, 2021MNRAS.502.2815C} is being carried out to develop parameter estimation techniques using 21~cm signal (both global and fluctuation) and reionization related observations. These parameter estimation techniques can broadly be classified into two categories - one is based on Bayesian method and another is based on machine learning techniques. Before the discovery of EDGES signal, \citet{2016MNRAS.457.1864L} have used a combination of mock 21 cm power spectrum, a mock global 21 cm signal (in the redshift range $z \sim 10-14$) and Planck observations found a much stringent constraints on cosmological and astrophysical parameters compared to Planck constraints.
More recently, \cite{2021MNRAS.507.2405C} with the help of an advanced MCMC based pipeline, \texttt{CosmoReionMC}, has used Planck measurements and QSO related observations along with the EDGES signal to put constraints jointly on the cosmological and astrophysical parameters by simultaneously varying all the free parameters. Their finding suggests that when EDGES signal is used along with the QSO related observations such as photoionization rate, the redshift distribution of Lyman limit system and neutral hydrogen fraction at $z\sim 5-6$, the introduction of early Pop~III stars becomes unavoidable. They also find that if the EDGES signal is replaced by a mock 21~cm signal generated from a standard galaxy formation code, the constraints on the cosmological parameters become much more stringent compared to the earlier constraints coming from only Planck observations \citep{Planck18}. On the other hand, \cite{2021MNRAS.502.2815C} had developed a artificial neural network based technique to extract astrophysical parameters from EDGES observation whereas \cite{2019MNRAS.484..282G}, used  a convolution neural network based pipeline to extract the astrophysical parameters directly from 21-cm images.

Besides the global signal, the power spectrum measurements have also been used to constrain the cosmic dawn and EoR. While these upper limits are becoming stronger with the improvement of data analysis techniques and adding more observation hours, they are still at least one order larger than the expected large-scale 21-cm power spectrum.  Most of the upper limits on the signal power spectrum obtained from different radio interferometric observations are unable to rule out cosmic dawn and EoR scenarios which do not require either an unconventional cooling mechanism or the presence of strong additional radio background to the CMB. Especially, these upper limits become weaker at a higher redshift.  Nevertheless, recent results from LOFAR, MWA, HERA started to rule out some models of EoR and cosmic dawn. For example, using the recent results from HERA which appeared in \cite{2022ApJ...925..221A}, the team showed that the IGM temperature must be larger than the adiabatic cooling threshold by redshift 8 while the soft band X-ray luminosities per star formation rate of the first galaxies are constrained ($1\sigma$ level) to [$10^{40.2}, 10^{41.9}$] erg/s/($\MSUN$/yr) \citep{2022ApJ...924...51A}. 
Besides constraining the properties of the early astrophysical sources and processes that have a significant impact on the 21-cm signal through heating/cooling of the IGM gas or by changing the radio background, one can also study the properties of the IGM. Recent studies such as \citet[][]{2020MNRAS.493.4728G, 2020arXiv200603203G, 2021MNRAS.503.4551G} used the results from LOFAR, MWA, and aimed to constrain the ionization and thermal states of the IGM in terms of quantities such as the average ionization fraction, average gas temperature, averaged brightness temperature, the volume fraction of the `heated regions’ IGM with temperature larger than $T_{\rm CMB}$, characteristic size of these hearts regions, etc.

One should realize that, although the power spectrum is very useful for characterising the spatial fluctuations of the desired signal, it is not able  to provide all spatial information hidden in the signal, as the 21-cm signal during cosmic dawn and EoR is expected to be highly non-gaussian. Thus, we need higher-order statistics such as bispectrum \citep[see e.g.,][]{majumdar18, 2019JCAP...02..058G, 2021MNRAS.502.3800K, 2021arXiv210808201K} to reveal such missing information which the power spectrum does not encode.


\section{HI 21-cm images during cosmic dawn}
\label{sec:image}
In general, one can expect four distinct features around a source during the cosmic dawn ( e.g., see the right panel of Fig.~\ref{image_tbprof}). These separate regions, from the centre, are: 
\begin{enumerate}
    \item HII/ionized region: UV photons from the source ionize the medium just adjacent to the centre. Consequently, both the HI fraction $x_{\rm HI}$ and the 21-cm signal become zero in that region. 
    \item Emission region: X-rays/cosmic rays etc. from the source penetrate longer distance than the UV photons due to their longer mean free path and heat the gas in the IGM. This creates a region outside the central HII region where the IGM kinetic temperature $T_{\rm K} \approx T_S > T_\gamma$. As a result the HI differential brightness temperature $\delta T_b$ becomes positive i.e., the 21-cm signal is seen in emission in that region. 
    \item Absorption region: Small amount of Ly-$\alpha$ photons is enough to couple the HI spin temperature $T_s$ with the IGM kinetic temperature $T_k$. Therefore, the above emission region is likely to be followed by a region where the Lyman-$\alpha$ coupling is strong but the X-ray/cosmic ray heating is not efficient. This makes $T_s < T_\gamma$ and $\delta T_b$ negative in that region i.e., the HI 21-cm signal will be seen in absorption.   
    \item Ly-$\alpha$ deficient region: The Ly-$\alpha$ coupling becomes inefficient at region far away from the source centre. In this far away region  $T_s=T_\gamma$ and thus the HI 21-cm signal vanishes there.
\end{enumerate}
Length scales of these distinct regions depend on the properties of the source. 
It is clear from the figure  (see left panel of Fig.~\ref{image_tbprof}) that the strength of the signal as well as the size of the emission region are much weaker than the absorption region for  typical early generation sources such as early galaxies, high mass X-ray binaries (HMXBs). We see that the HI 21-cm signal profile around an isolated source is very simple. However, in reality, one expects significant overlap between these individual 21 cm patterns as there will be multiple sources in the IGM. Nevertheless, one might expect isolated absorption regions around clusters of sources at the initial stages of the cosmic dawn when the Ly-$\alpha$ coupling yet to become saturated. This era is expected to be followed by an epoch of IGM heating. Ly-$\alpha$ coupling in this era is expected to be strong.  Fig.~\ref{image_tbmap} shows a simulated 2D differential brightness temperature map of such an epoch where large emission regions (in yellow) are embedded in absorption background (in green) (see Fig.~\ref{image_tbmap}). At the end of this heating era, the IGM kinetic and spin temperature are expected to become higher than the CMBR temperature i.e.,  $T_s \approx T_k >>T_\gamma$.  One also expects ionized regions of different sizes grow with time and get overlapped with each other during this era. 

Fig.~\ref{fig:image_pszk} shows large scale HI 21-cm power spectrum as a function of redshift. We see that there are three `peaks' appearing from higher to lower redshifts  corresponding to fluctuations in the  Ly-$\alpha$ coupling, X-ray heating and ionization field respectively.  Here we note that there might be the scenarios where there could be overlap between Ly-$\alpha$ coupling, heating and ionization processes, and these three distinct peak feature may not arise there.

Detection of the cosmological HI 21-cm signal is quite challenging as it is very weak compared to the astrophysical foregrounds and the system noise.  Due to limited sensitivity, the existing radio interferometers (e.g., uGMRT, LOFAR, MWA etc. ) mainly aim to measure the signal statistically using power spectrum, Bi-spectrum etc. However, it has been proposed and studied in detail \citep{datta07b, datta08, majumdar11, datta12a} that a matched filter based technique can be sued to detect individual HI/HII regions using ongoing /upcoming experiments. Some of the ongoing interferometric experiments can also produce low-resolution images of the EoR and cosmic dawn 21-cm signal \citep[see e.g.,][]{2012MNRAS.425.2964Z}. On the contrary, the upcoming Square Kilometre Array (SKA) will have adequate sensitivity to produce actual tomographic images of the HI signal with good enough resolutions \citep{2015aska.confE..10M, ghara16}. Motivated by the above prospect, several theoretical studies have started proposing methods for extracting information about the cosmic dawn and EoR using tomographic analysis of HI images. Currently, it is not possible to do pixel to pixel comparison between an observed tomographic image and a model image of the 21-cm signal.  Thus, these studies try to statistically characterise the topology of the images.  The use of Minkowski functionals \citep[e.g.,][]{ 2018MNRAS.477.1984B, 2019MNRAS.485.2235B,2021JCAP...05..026K}, Euler characteristic \citep[see e.g.,][]{2021MNRAS.505.1863G}, Fractal dimensions \citep[e.g.,][]{2017MNRAS.466.2302B}, Bubble size distributions \citep[][]{2018MNRAS.479.5596G, 2020MNRAS.496..739G}, Individual images using convolutional neural network \citep[e.g.,][]{2019MNRAS.484..282G} are some such approaches.


\section{Observations with SKA and synergies with global signal}
\label{sec:synergy}
Measurements of the global signal and statistical quantities such as the power spectra/Bi-spectra of redshifted HI 21-cm radiation are two major observational strategies for probing the cosmic dawn and reionization era. Highly sensitive experiments such as the SKA and HERA should also be able to detect images of HI 21-cm fields.  HI 21-cm signal measured using different strategies encodes signatures of the same underlying sources, inter-galactic medium and physical processes. Therefore, a joint analysis of these signals is expected to constrain models of cosmic dawn much better. 

In particular, the global HI 21-cm signal carries information about the redshift evolution of various radiation fields such as Ly-$\alpha$, X-rays, UV, Lyman-Werner, radio  and physical processes such as IGM heating due to X-rays/cosmic rays/magnetic fields and Wouthuysen-Field coupling etc. The global signal is the strongest at higher redshift when the Ly-$\alpha$ coupling is saturated and the IGM heating is not so significant (refer to the top panel of Fig.~\ref{fig:image_pszk}) \citep{pritchard12}. It also peaks at lower redshift when the IGM heating is substantial and spin temperature is much higher than the CMBR temperature (Fig.~\ref{fig:image_pszk}). On the other hand, the HI 21-cm power spectrum  peaks at three different stages corresponding to maximum fluctuations due to Ly-$\alpha$ coupling, IGM heating and ionization during the cosmic dawn and reionization era \citep{baek10, ghara15a}. It is observed that the power spectra peak when the coupling, heating and ionization are roughly at half-way respectively (refer to the middle panel of Fig.~\ref{fig:image_pszk}).  It is also seen that there are two dips in the power spectrum-redshift plot. The one at higher redshift corresponds to  the situation when the spin temperature becomes fully coupled to the IGM temperature. The dip at lower redshift arises when IGM temperature is much higher than the CMBR temperature but ionization is at initial stage. 

We see from the above discussion that information contained in the global signal and power spectrum can be treated as complimentary to each other in a sense that the global signal is maximum when the power spectrum is minimum and vice versa. Therefore, a joint analysis of these two signals would be very useful for understanding the cosmic dawn much better. A thorough analysis is required towards this. Here we note that there is a separate article which reviews synergies between cosmological 21-cm signal and various line intensity maps in order to probe the EoR \citep{Murmu2022joaa}.  

\section{Summary}
\label{sec:summary}
Observations of the cosmological HI 21-cm signal using the Square Kilometer array (SKA) are expected to revolutionize our understanding about the early Universe, particularly, the period when the first luminous sources appeared in the Universe. Here, we review recent theoretical developments regarding the redshifted HI 21-cm signal from cosmic dawn, prospects of constraining first luminous sources and inter-galactic medium using ongoing and upcoming experiments.  A companion review article \citep{skamo2} focuses on the HI 21-cm signal from the epoch of reionization and various observational challenges for detecting the signal.

We begin with a short description of basics of cosmological HI 21-cm signal with emphasis on two main observable quantities i.e., the global HI 21-cm signal and the power spectrum of HI brightness temperature fluctuations. The first generation of luminous sources play an important role in shaping both the global signal and power spectrum. Therefore, we present a discussion on the possible first sources such as Pop~III, Pop~II stars, galaxies, mini-quasars that are likely to influence the HI 21-cm signal during the cosmic dawn. Subsequently, radiation background produced by these sources at different wavebands such as the X-ray, Ly-$\alpha$, Lyman-Werner, (excess) radio radiation during the cosmic dawn are discussed. In particular, we focus on the redshift evolution of these radiation background and subsequent feedback effects on the source formation. Additionally, we  discuss production of cosmic rays from Pop~III and Pop~II sources and their impact on the thermal state of the IGM during cosmic dawn. 

We, then, highlight various physical processes such as the coupling of the spin temperature with the IGM kinetic temperature through the Wouthuysen-Field effects, heating of the IGM and increase of background temperature corresponding to radio radiation that are very important. Several mechanisms of IGM heating/cooling by soft X-rays from stars, mini-quasars, the primordial magnetic field, cosmic rays, dark matter baryonic interaction have been discussed . Impacts of all these on the redshift evolution of the spin and IGM temperature have also been discussed. Finally, we discuss their impacts on the global HI 21-cm signal.

Other major topics considered in this review are the HI 21-cm power spectrum and prospects of imaging the HI 21-cm field using the SKA. Here, we mostly focus on results from numerical simulations and discuss unique features in the redshift evolution of large scale HI 21-cm power spectrum. We put a special emphasis on the `three peak' nature of the HI power spectrum when plotted against redshift.

Further, we summarize works which study constraints on the models of HI 21-cm signal during cosmic dawn and reionization using existing measurements of the global 21-cm signal and upper limits on the power spectrum obtained from ongoing interferometers such as the GMRT, MWA, HERA, LOFAR. Finally, we discuss possibilities of constraining the cosmic dawn using a joint analysis of  the global HI 21-cm signal and power spectrum measured by interoferometric experiments such the SKA.



\section*{Acknowledgements}
AB acknowledges financial support from UGC, Govt. of India. SS and KKD acknowledge financial support from BRNS through a project grant (sanction no: 57/14/10/2019-BRNS). SS thanks Presidency University for the support through FRPDF grant.
RG acknowledges support by the Israel Science Foundation (grant no. 255/18). AC acknowledge support of the Department of Atomic Energy, Government of India, under project no.~12-R\&D-TFR-5.02-0700. KKD also acknowledges financial support from SERB-DST (Govt. of India) through a project under MATRICS scheme (MTR/2021/000384).

\vspace{-1em}


\bibliography{ref}

\begin{thebibliography}{}
\expandafter\ifx\csname natexlab\endcsname\relax\def\natexlab#1{#1}\fi

\bibitem[{{Abdurashidova} {$et~al$.}(2022{\natexlab{a}}){Abdurashidova},
  {Aguirre}, {Alexander}, {Ali}, {Balfour}, {Beardsley}, {Bernardi},
  {Billings}, {Bowman}, {Bradley}, {Bull}, {Burba}, {Carey}, {Carilli},
  {Cheng}, {DeBoer}, {Dexter}, {de Lera Acedo}, {Dibblee-Barkman}, {Dillon},
  {Ely}, {Ewall-Wice}, {Fagnoni}, {Fritz}, {Furlanetto}, {Gale-Sides},
  {Glendenning}, {Gorthi}, {Greig}, {Grobbelaar}, {Halday}, {Hazelton},
  {Hewitt}, {Hickish}, {Jacobs}, {Julius}, {Kern}, {Kerrigan}, {Kittiwisit},
  {Kohn}, {Kolopanis}, {Lanman}, {La Plante}, {Lekalake}, {Lewis}, {Liu},
  {MacMahon}, {Malan}, {Malgas}, {Maree}, {Martinot}, {Matsetela}, {Mesinger},
  {Molewa}, {Morales}, {Mosiane}, {Murray}, {Neben}, {Nikolic}, {Nunhokee},
  {Parsons}, {Patra}, {Pascua}, {Pieterse}, {Pober}, {Razavi-Ghods},
  {Ringuette}, {Robnett}, {Rosie}, {Sims}, {Singh}, {Smith}, {Syce},
  {Thyagarajan}, {Williams}, {Zheng}, \& {HERA
  Collaboration}}]{2022ApJ...925..221A}
{Abdurashidova}, Z., {Aguirre}, J.~E., {Alexander}, P., {$et~al$.}
  2022{\natexlab{a}}, \apj, 925, 221

\bibitem[{{Abdurashidova} {$et~al$.}(2022{\natexlab{b}}){Abdurashidova},
  {Aguirre}, {Alexander}, {Ali}, {Balfour}, {Barkana}, {Beardsley}, {Bernardi},
  {Billings}, {Bowman}, {Bradley}, {Bull}, {Burba}, {Carey}, {Carilli},
  {Cheng}, {DeBoer}, {Dexter}, {de Lera Acedo}, {Dillon}, {Ely}, {Ewall-Wice},
  {Fagnoni}, {Fialkov}, {Fritz}, {Furlanetto}, {Gale-Sides}, {Glendenning},
  {Gorthi}, {Greig}, {Grobbelaar}, {Halday}, {Hazelton}, {Heimersheim},
  {Hewitt}, {Hickish}, {Jacobs}, {Julius}, {Kern}, {Kerrigan}, {Kittiwisit},
  {Kohn}, {Kolopanis}, {Lanman}, {La Plante}, {Lekalake}, {Lewis}, {Liu}, {Ma},
  {MacMahon}, {Malan}, {Malgas}, {Maree}, {Martinot}, {Matsetela}, {Mesinger},
  {Mirocha}, {Molewa}, {Morales}, {Mosiane}, {Mu{\~n}oz}, {Murray}, {Neben},
  {Nikolic}, {Nunhokee}, {Parsons}, {Patra}, {Pieterse}, {Pober}, {Qin},
  {Razavi-Ghods}, {Reis}, {Ringuette}, {Robnett}, {Rosie}, {Santos}, {Sikder},
  {Sims}, {Smith}, {Syce}, {Thyagarajan}, {Williams}, \& {Zheng}}]{hera22}
---. 2022{\natexlab{b}}, \apj, 924, 51

\bibitem[{{Abdurashidova} {$et~al$.}(2022{\natexlab{c}}){Abdurashidova},
  {Aguirre}, {Alexander}, {Ali}, {Balfour}, {Barkana}, {Beardsley}, {Bernardi},
  {Billings}, {Bowman}, {Bradley}, {Bull}, {Burba}, {Carey}, {Carilli},
  {Cheng}, {DeBoer}, {Dexter}, {de Lera Acedo}, {Dillon}, {Ely}, {Ewall-Wice},
  {Fagnoni}, {Fialkov}, {Fritz}, {Furlanetto}, {Gale-Sides}, {Glendenning},
  {Gorthi}, {Greig}, {Grobbelaar}, {Halday}, {Hazelton}, {Heimersheim},
  {Hewitt}, {Hickish}, {Jacobs}, {Julius}, {Kern}, {Kerrigan}, {Kittiwisit},
  {Kohn}, {Kolopanis}, {Lanman}, {La Plante}, {Lekalake}, {Lewis}, {Liu}, {Ma},
  {MacMahon}, {Malan}, {Malgas}, {Maree}, {Martinot}, {Matsetela}, {Mesinger},
  {Mirocha}, {Molewa}, {Morales}, {Mosiane}, {Mu{\~n}oz}, {Murray}, {Neben},
  {Nikolic}, {Nunhokee}, {Parsons}, {Patra}, {Pieterse}, {Pober}, {Qin},
  {Razavi-Ghods}, {Reis}, {Ringuette}, {Robnett}, {Rosie}, {Santos}, {Sikder},
  {Sims}, {Smith}, {Syce}, {Thyagarajan}, {Williams}, \&
  {Zheng}}]{2022ApJ...924...51A}
---. 2022{\natexlab{c}}, \apj, 924, 51

\bibitem[{{Abe} \& {Tashiro}(2021)}]{abe2021}
{Abe}, K.~T., \& {Tashiro}, H. 2021, \prd, 103, 123543

\bibitem[{{Abel} {$et~al$.}(2002){Abel}, {Bryan}, \& {Norman}}]{abel02}
{Abel}, T., {Bryan}, G.~L., \& {Norman}, M.~L. 2002, Science, 295, 93

\bibitem[{{Ali} {$et~al$.}(2008){Ali}, {Bharadwaj}, \& {Chengalur}}]{Ali_2008}
{Ali}, S.~S., {Bharadwaj}, S., \& {Chengalur}, J.~N. 2008, \mnras, 385, 2166

\bibitem[{{Baek} {$et~al$.}(2010){Baek}, {Semelin}, {Di Matteo}, {Revaz}, \&
  {Combes}}]{baek10}
{Baek}, S., {Semelin}, B., {Di Matteo}, P., {Revaz}, Y., \& {Combes}, F. 2010,
  \aap, 523, A4

\bibitem[{{Bag} {$et~al$.}(2019){Bag}, {Mondal}, {Sarkar}, {Bharadwaj},
  {Choudhury}, \& {Sahni}}]{2019MNRAS.485.2235B}
{Bag}, S., {Mondal}, R., {Sarkar}, P., {$et~al$.} 2019, \mnras, 485, 2235

\bibitem[{{Bag} {$et~al$.}(2018){Bag}, {Mondal}, {Sarkar}, {Bharadwaj}, \&
  {Sahni}}]{2018MNRAS.477.1984B}
{Bag}, S., {Mondal}, R., {Sarkar}, P., {Bharadwaj}, S., \& {Sahni}, V. 2018,
  \mnras, 477, 1984

\bibitem[{{Bandyopadhyay} {$et~al$.}(2017){Bandyopadhyay}, {Choudhury}, \&
  {Seshadri}}]{2017MNRAS.466.2302B}
{Bandyopadhyay}, B., {Choudhury}, T.~R., \& {Seshadri}, T.~R. 2017, \mnras,
  466, 2302

\bibitem[{{Barkana}(2018)}]{Barkana18Nature}
{Barkana}, R. 2018, \nat, 555, 71

\bibitem[{{Barkana} \& {Loeb}(2001)}]{barkana01}
{Barkana}, R., \& {Loeb}, A. 2001, Physics Reports, 349, 125

\bibitem[{Barkana \& Loeb(2005)}]{Barkana_2005}
Barkana, R., \& Loeb, A. 2005, The Astrophysical Journal, 626, 1

\bibitem[{{Barry} {$et~al$.}(2019){Barry}, {Wilensky}, {Trott}, {Pindor},
  {Beardsley}, {Hazelton}, {Sullivan}, {Morales}, {Pober}, {Line}, {Greig},
  {Byrne}, {Lanman}, {Li}, {Jordan}, {Joseph}, {McKinley}, {Rahimi},
  {Yoshiura}, {Bowman}, {Gaensler}, {Hewitt}, {Jacobs}, {Mitchell}, {Udaya
  Shankar}, {Sethi}, {Subrahmanyan}, {Tingay}, {Webster}, \&
  {Wyithe}}]{2019ApJ...884....1B}
{Barry}, N., {Wilensky}, M., {Trott}, C.~M., {$et~al$.} 2019, \apj, 884, 1

\bibitem[{{Beardsley} {$et~al$.}(2016){Beardsley}, {Hazelton}, {Sullivan},
  {Carroll}, {Barry}, {Rahimi}, {Pindor}, {Trott}, {Line}, {Jacobs}, {Morales},
  {Pober}, {Bernardi}, {Bowman}, {Busch}, {Briggs}, {Cappallo}, {Corey}, {de
  Oliveira-Costa}, {Dillon}, {Emrich}, {Ewall-Wice}, {Feng}, {Gaensler},
  {Goeke}, {Greenhill}, {Hewitt}, {Hurley-Walker}, {Johnston-Hollitt},
  {Kaplan}, {Kasper}, {Kim}, {Kratzenberg}, {Lenc}, {Loeb}, {Lonsdale},
  {Lynch}, {McKinley}, {McWhirter}, {Mitchell}, {Morgan}, {Neben},
  {Thyagarajan}, {Oberoi}, {Offringa}, {Ord}, {Paul}, {Prabu}, {Procopio},
  {Riding}, {Rogers}, {Roshi}, {Udaya Shankar}, {Sethi}, {Srivani},
  {Subrahmanyan}, {Tegmark}, {Tingay}, {Waterson}, {Wayth}, {Webster},
  {Whitney}, {Williams}, {Williams}, {Wu}, \& {Wyithe}}]{2016ApJ...833..102B}
{Beardsley}, A.~P., {Hazelton}, B.~J., {Sullivan}, I.~S., {$et~al$.} 2016,
  \apj, 833, 102

\bibitem[{{Bell}(1978)}]{Bell1978}
{Bell}, A.~R. 1978, \mnras, 182, 147

\bibitem[{{Benson}(2012)}]{Galacticus}
{Benson}, A.~J. 2012, \na, 17, 175

\bibitem[{{Bera} {$et~al$.}(2020){Bera}, {Datta}, \& {Samui}}]{Bera_2020}
{Bera}, A., {Datta}, K.~K., \& {Samui}, S. 2020, \mnras, 498, 918

\bibitem[{{Bera} {$et~al$.}(2022){Bera}, {Samui}, \& {Datta}}]{Bera2022}
{Bera}, A., {Samui}, S., \& {Datta}, K.~K. 2022, arXiv e-prints,
  arXiv:2202.12308

\bibitem[{Berlin {$et~al$.}(2018)Berlin, Hooper, Krnjaic, \&
  McDermott}]{PhysRevLett.121.011102}
Berlin, A., Hooper, D., Krnjaic, G., \& McDermott, S.~D. 2018, Phys. Rev.
  Lett., 121, 011102

\bibitem[{{Bharadwaj} \& {Ali}(2004)}]{Bharadwaj_2004}
{Bharadwaj}, S., \& {Ali}, S.~S. 2004, \mnras, 352, 142

\bibitem[{{Bharadwaj} \& {Ali}(2005)}]{Bharadwaj_2005}
---. 2005, \mnras, 356, 1519

\bibitem[{{Bhatt, Jitesh R.} {$et~al$.}(2020){Bhatt, Jitesh R.}, {Natwariya,
  Pravin Kumar}, {Nayak, Alekha C.}, \& {Pandey, Arun Kumar}}]{bhatt2020}
{Bhatt, Jitesh R.}, {Natwariya, Pravin Kumar}, {Nayak, Alekha C.}, \& {Pandey,
  Arun Kumar}. 2020, Eur. Phys. J. C, 80, 334

\bibitem[{{Bowman} {$et~al$.}(2018{\natexlab{a}}){Bowman}, {Rogers},
  {Monsalve}, {Mozdzen}, \& {Mahesh}}]{EDGES18}
{Bowman}, J.~D., {Rogers}, A. E.~E., {Monsalve}, R.~A., {Mozdzen}, T.~J., \&
  {Mahesh}, N. 2018{\natexlab{a}}, \nat, 555, 67

\bibitem[{{Bowman} {$et~al$.}(2018{\natexlab{b}}){Bowman}, {Rogers},
  {Monsalve}, {Mozdzen}, \& {Mahesh}}]{EDGES2018}
{Bowman}, J.~D., {Rogers}, A.~E.~E., {Monsalve}, R.~A., {Mozdzen}, T.~J., \&
  {Mahesh}, N. 2018{\natexlab{b}}, \nat, 555, 67

\bibitem[{Bradley {$et~al$.}(2019)Bradley, Tauscher, Rapetti, \&
  Burns}]{Bradley_2019}
Bradley, R.~F., Tauscher, K., Rapetti, D., \& Burns, J.~O. 2019, The
  Astrophysical Journal, 874, 153

\bibitem[{{Bromm} {$et~al$.}(1999){Bromm}, {Coppi}, \& {Larson}}]{bromm99}
{Bromm}, V., {Coppi}, P.~S., \& {Larson}, R.~B. 1999, \apjl, 527, L5

\bibitem[{{Bromm} \& {Larson}(2004)}]{bromm04}
{Bromm}, V., \& {Larson}, R.~B. 2004, \araa, 42, 79

\bibitem[{{Chakraborty} {$et~al$.}(2019){Chakraborty}, {Datta}, {Choudhuri},
  {Roy}, {Intema}, {Choudhury}, {Datta}, {Pal}, {Bharadwaj}, {Dutta}, \&
  {Choudhury}}]{Arnab_GSB_2019a}
{Chakraborty}, A., {Datta}, A., {Choudhuri}, S., {$et~al$.} 2019, \mnras, 487,
  4102

\bibitem[{{Chatterjee} {$et~al$.}(2021){Chatterjee}, {Choudhury}, \&
  {Mitra}}]{2021MNRAS.507.2405C}
{Chatterjee}, A., {Choudhury}, T.~R., \& {Mitra}, S. 2021, \mnras, 507, 2405

\bibitem[{{Chatterjee} {$et~al$.}(2020){Chatterjee}, {Dayal}, {Choudhury}, \&
  {Schneider}}]{atri20}
{Chatterjee}, A., {Dayal}, P., {Choudhury}, T.~R., \& {Schneider}, R. 2020,
  \mnras, 496, 1445

\bibitem[{{Chen} \& {Miralda-Escud{\'e}}(2004)}]{Chen_2004}
{Chen}, X., \& {Miralda-Escud{\'e}}, J. 2004, \apj, 602, 1

\bibitem[{{Chluba} {$et~al$.}(2015){Chluba}, {Paoletti}, {Finelli}, \&
  {Rubi{\~n}o-Mart{\'\i}n}}]{Chluba15}
{Chluba}, J., {Paoletti}, D., {Finelli}, F., \& {Rubi{\~n}o-Mart{\'\i}n}, J.~A.
  2015, \mnras, 451, 2244

\bibitem[{{Choudhury} {$et~al$.}(2021){Choudhury}, {Chatterjee}, {Datta}, \&
  {Choudhury}}]{2021MNRAS.502.2815C}
{Choudhury}, M., {Chatterjee}, A., {Datta}, A., \& {Choudhury}, T.~R. 2021,
  \mnras, 502, 2815

\bibitem[{{Choudhury} \& {Ferrara}(2005)}]{2005MNRAS.361..577C}
{Choudhury}, T.~R., \& {Ferrara}, A. 2005, \mnras, 361, 577

\bibitem[{{Choudhury} \& {Ferrara}(2006)}]{2006MNRAS.371L..55C}
---. 2006, \mnras, 371, L55

\bibitem[{{Cole} {$et~al$.}(2000){Cole}, {Lacey}, {Baugh}, \&
  {Frenk}}]{Galform}
{Cole}, S., {Lacey}, C.~G., {Baugh}, C.~M., \& {Frenk}, C.~S. 2000, \mnras,
  319, 168

\bibitem[{{Cooray} {$et~al$.}(2019){Cooray}, {Aguirre}, {Ali-Haimoud},
  {Alvarez}, {Appleton}, {Armus}, {Becker}, {Bock}, {Bowler}, {Bowman},
  {Bradford}, {Breysse}, {Bromm}, {Burns}, {Caputi}, {Castellano}, {Chang},
  {Chary}, {Chiang}, {Cohn}, {Conselice}, {Cuby}, {Davies}, {Dayal}, {Dore},
  {Farrah}, {Ferrara}, {Finkelstein}, {Furlanetto}, {Hazelton}, {Heneka},
  {Hutter}, {Jacobs}, {Koopmans}, {Kovetz}, {La Piante}, {Le Fevre}, {Liu},
  {Ma}, {Ma}, {Malhotra}, {Mao}, {Marrone}, {Masui}, {McQuinn}, {Mirocha},
  {Mortlock}, {Murphy}, {Nayyeri}, {Natarajan}, {Nithyanandan}, {Parsons},
  {Pello}, {Pope}, {Rhoads}, {Rhodes}, {Riechers}, {Robertson}, {Scarlata},
  {Serjeant}, {Saliwanchik}, {Salvaterra}, {Schneider}, {Silva}, {Sahl{\'e}n},
  {Santos}, {Switzer}, {Temi}, {Trac}, {Venkatesan}, {Visbal}, {Zaldarriaga},
  {Zemcov}, \& {Zheng}}]{2019BAAS...51c..48C}
{Cooray}, A., {Aguirre}, J., {Ali-Haimoud}, Y., {$et~al$.} 2019, \baas, 51, 48

\bibitem[{{Datta} {$et~al$.}(2007){Datta}, {Bharadwaj}, \&
  {Choudhury}}]{datta07b}
{Datta}, K.~K., {Bharadwaj}, S., \& {Choudhury}, T.~R. 2007, \mnras, 382, 809

\bibitem[{{Datta} {$et~al$.}(2012{\natexlab{a}}){Datta}, {Friedrich},
  {Mellema}, {Iliev}, \& {Shapiro}}]{datta12a}
{Datta}, K.~K., {Friedrich}, M.~M., {Mellema}, G., {Iliev}, I.~T., \&
  {Shapiro}, P.~R. 2012{\natexlab{a}}, \mnras, 424, 762

\bibitem[{{Datta} {$et~al$.}(2014){Datta}, {Jensen}, {Majumdar}, {Mellema},
  {Iliev}, {Mao}, {Shapiro}, \& {Ahn}}]{datta14}
{Datta}, K.~K., {Jensen}, H., {Majumdar}, S., {$et~al$.} 2014, \mnras, 442,
  1491

\bibitem[{{Datta} {$et~al$.}(2008){Datta}, {Majumdar}, {Bharadwaj}, \&
  {Choudhury}}]{datta08}
{Datta}, K.~K., {Majumdar}, S., {Bharadwaj}, S., \& {Choudhury}, T.~R. 2008,
  \mnras, 391, 1900

\bibitem[{{Datta} {$et~al$.}(2012{\natexlab{b}}){Datta}, {Mellema}, {Mao},
  {Iliev}, {Shapiro}, \& {Ahn}}]{Datta2012b}
{Datta}, K.~K., {Mellema}, G., {Mao}, Y., {$et~al$.} 2012{\natexlab{b}},
  \mnras, 424, 1877

\bibitem[{{Dayal} \& {Ferrara}(2018)}]{dayal2018}
{Dayal}, P., \& {Ferrara}, A. 2018, \physrep, 780, 1

\bibitem[{De \& Acedo(2019)}]{unknown}
De, E., \& Acedo, E. 2019, REACH: Radio Experiment for the Analysis of Cosmic
  Hydrogen

\bibitem[{{DeBoer} {$et~al$.}(2017){DeBoer}, {Parsons}, {Aguirre}, {Alexander},
  {Ali}, {Beardsley}, {Bernardi}, {Bowman}, {Bradley}, {Carilli}, {Cheng}, {de
  Lera Acedo}, {Dillon}, {Ewall-Wice}, {Fadana}, {Fagnoni}, {Fritz},
  {Furlanetto}, {Glendenning}, {Greig}, {Grobbelaar}, {Hazelton}, {Hewitt},
  {Hickish}, {Jacobs}, {Julius}, {Kariseb}, {Kohn}, {Lekalake}, {Liu}, {Loots},
  {MacMahon}, {Malan}, {Malgas}, {Maree}, {Martinot}, {Mathison}, {Matsetela},
  {Mesinger}, {Morales}, {Neben}, {Patra}, {Pieterse}, {Pober}, {Razavi-Ghods},
  {Ringuette}, {Robnett}, {Rosie}, {Sell}, {Smith}, {Syce}, {Tegmark},
  {Thyagarajan}, {Williams}, \& {Zheng}}]{2017PASP..129d5001D}
{DeBoer}, D.~R., {Parsons}, A.~R., {Aguirre}, J.~E., {$et~al$.} 2017, \pasp,
  129, 045001

\bibitem[{{Dowell} \& {Taylor}(2018)}]{2018ApJ...858L...9D}
{Dowell}, J., \& {Taylor}, G.~B. 2018, \apjl, 858, L9

\bibitem[{{Eastwood} {$et~al$.}(2019){Eastwood}, {Anderson}, {Monroe},
  {Hallinan}, {Catha}, {Dowell}, {Garsden}, {Greenhill}, {Hicks}, {Kocz},
  {Price}, {Schinzel}, {Vedantham}, \& {Wang}}]{2019AJ....158...84E}
{Eastwood}, M.~W., {Anderson}, M.~M., {Monroe}, R.~M., {$et~al$.} 2019, \aj,
  158, 84

\bibitem[{{Ewall-Wice} {$et~al$.}(2018){Ewall-Wice}, {Chang}, {Lazio},
  {Dor{\'e}}, {Seiffert}, \& {Monsalve}}]{Ewall_2018}
{Ewall-Wice}, A., {Chang}, T.~C., {Lazio}, J., {$et~al$.} 2018, \apj, 868, 63

\bibitem[{{Ewall-Wice} {$et~al$.}(2020){Ewall-Wice}, {Chang}, \&
  {Lazio}}]{2020MNRAS.492.6086E}
{Ewall-Wice}, A., {Chang}, T.-C., \& {Lazio}, T. J.~W. 2020, \mnras, 492, 6086

\bibitem[{{Ewall-Wice} {$et~al$.}(2016){Ewall-Wice}, {Dillon}, {Hewitt},
  {Loeb}, {Mesinger}, {Neben}, {Offringa}, {Tegmark}, {Barry}, {Beardsley},
  {Bernardi}, {Bowman}, {Briggs}, {Cappallo}, {Carroll}, {Corey}, {de
  Oliveira-Costa}, {Emrich}, {Feng}, {Gaensler}, {Goeke}, {Greenhill},
  {Hazelton}, {Hurley-Walker}, {Johnston-Hollitt}, {Jacobs}, {Kaplan},
  {Kasper}, {Kim}, {Kratzenberg}, {Lenc}, {Line}, {Lonsdale}, {Lynch},
  {McKinley}, {McWhirter}, {Mitchell}, {Morales}, {Morgan}, {Thyagarajan},
  {Oberoi}, {Ord}, {Paul}, {Pindor}, {Pober}, {Prabu}, {Procopio}, {Riding},
  {Rogers}, {Roshi}, {Shankar}, {Sethi}, {Srivani}, {Subrahmanyan}, {Sullivan},
  {Tingay}, {Trott}, {Waterson}, {Wayth}, {Webster}, {Whitney}, {Williams},
  {Williams}, {Wu}, \& {Wyithe}}]{2016MNRAS.460.4320E}
{Ewall-Wice}, A., {Dillon}, J.~S., {Hewitt}, J.~N., {$et~al$.} 2016, \mnras,
  460, 4320

\bibitem[{{Feng} \& {Holder}(2018{\natexlab{a}})}]{2018ApJ...858L..17F}
{Feng}, C., \& {Holder}, G. 2018{\natexlab{a}}, \apjl, 858, L17

\bibitem[{{Feng} \& {Holder}(2018{\natexlab{b}})}]{Feng_2018}
---. 2018{\natexlab{b}}, \apjl, 858, L17

\bibitem[{{Fialkov} \& {Barkana}(2014)}]{fialkov14}
{Fialkov}, A., \& {Barkana}, R. 2014, \mnras, 445, 213

\bibitem[{{Fialkov} \& {Barkana}(2019)}]{Fialkov_2019}
---. 2019, \mnras, 486, 1763

\bibitem[{{Fialkov} {$et~al$.}(2018){Fialkov}, {Barkana}, \&
  {Cohen}}]{2018PhRvL.121a1101F}
{Fialkov}, A., {Barkana}, R., \& {Cohen}, A. 2018, \prl, 121, 011101

\bibitem[{{Field}(1958)}]{Field_1958}
{Field}, G.~B. 1958, Proceedings of the IRE, 46, 240

\bibitem[{{Fioc} \& {Rocca-Volmerange}(1997)}]{Fioc97}
{Fioc}, M., \& {Rocca-Volmerange}, B. 1997, \aap, 326, 950

\bibitem[{{Fixsen} {$et~al$.}(2011){Fixsen}, {Kogut}, {Levin}, {Limon},
  {Lubin}, {Mirel}, {Seiffert}, {Singal}, {Wollack}, {Villela}, \&
  {Wuensche}}]{2011ApJ...734....5F}
{Fixsen}, D.~J., {Kogut}, A., {Levin}, S., {$et~al$.} 2011, \apj, 734, 5

\bibitem[{{Furlanetto}(2006)}]{2006MNRAS.371..867F}
{Furlanetto}, S.~R. 2006, \mnras, 371, 867

\bibitem[{{Furlanetto} {$et~al$.}(2006{\natexlab{a}}){Furlanetto}, {Oh}, \&
  {Briggs}}]{furlanetto06}
{Furlanetto}, S.~R., {Oh}, S.~P., \& {Briggs}, F.~H. 2006{\natexlab{a}},
  Physics Reports, 433, 181

\bibitem[{{Furlanetto} {$et~al$.}(2006{\natexlab{b}}){Furlanetto}, {Oh}, \&
  {Pierpaoli}}]{furlanetto2006}
{Furlanetto}, S.~R., {Oh}, S.~P., \& {Pierpaoli}, E. 2006{\natexlab{b}}, \prd,
  74, 103502

\bibitem[{{Furlanetto} \& {Pritchard}(2006)}]{Furlanetto06MNRAS}
{Furlanetto}, S.~R., \& {Pritchard}, J.~R. 2006, \mnras, 372, 1093

\bibitem[{{Furlanetto} {$et~al$.}(2004){Furlanetto}, {Sokasian}, \&
  {Hernquist}}]{furlanetto04c}
{Furlanetto}, S.~R., {Sokasian}, A., \& {Hernquist}, L. 2004, \mnras, 347, 187

\bibitem[{{Gehlot} {$et~al$.}(2019){Gehlot}, {Mertens}, {Koopmans},
  {Brentjens}, {Zaroubi}, {Ciardi}, {Ghosh}, {Hatef}, {Iliev}, {Jeli{\'c}}, {},
  {Kooistra}, {Krause}, {Mellema}, {Mevius}, {Mitra}, {Offringa}, {Pandey},
  {Sardarabadi}, {Schaye}, {Silva}, {Vedantham}, \&
  {Yatawatta}}]{2019MNRAS.488.4271G}
{Gehlot}, B.~K., {Mertens}, F.~G., {Koopmans}, L.~V.~E., {$et~al$.} 2019,
  \mnras, 488, 4271

\bibitem[{{Gehlot} {$et~al$.}(2020){Gehlot}, {Mertens}, {Koopmans}, {Offringa},
  {Shulevski}, {Mevius}, {Brentjens}, {Kuiack}, {Pandey}, {Rowlinson},
  {Sardarabadi}, {Vedantham}, {Wijers}, {Yatawatta}, \&
  {Zaroubi}}]{2020MNRAS.499.4158G}
---. 2020, \mnras, 499, 4158

\bibitem[{{Gessey-Jones} {$et~al$.}(2022){Gessey-Jones}, {Sartorio}, {Fialkov},
  {Mirouh}, {Magg}, {de Lera Acedo}, {Handley}, \& {Barkana}}]{Gessey2022}
{Gessey-Jones}, T., {Sartorio}, N.~S., {Fialkov}, A., {$et~al$.} 2022, arXiv
  e-prints, arXiv:2202.02099

\bibitem[{{Ghara} \& {Choudhury}(2020)}]{2020MNRAS.496..739G}
{Ghara}, R., \& {Choudhury}, T.~R. 2020, \mnras, 496, 739

\bibitem[{{Ghara} {$et~al$.}(2015{\natexlab{a}}){Ghara}, {Choudhury}, \&
  {Datta}}]{ghara15a}
{Ghara}, R., {Choudhury}, T.~R., \& {Datta}, K.~K. 2015{\natexlab{a}}, \mnras,
  447, 1806

\bibitem[{{Ghara} {$et~al$.}(2016){Ghara}, {Choudhury}, \& {Datta}}]{ghara15c}
---. 2016, \mnras, 460, 827

\bibitem[{{Ghara} {$et~al$.}(2017){Ghara}, {Choudhury}, {Datta}, \&
  {Choudhuri}}]{ghara16}
{Ghara}, R., {Choudhury}, T.~R., {Datta}, K.~K., \& {Choudhuri}, S. 2017,
  \mnras, 464, 2234

\bibitem[{{Ghara} {$et~al$.}(2015{\natexlab{b}}){Ghara}, {Datta}, \&
  {Choudhury}}]{ghara15b}
{Ghara}, R., {Datta}, K.~K., \& {Choudhury}, T.~R. 2015{\natexlab{b}}, \mnras,
  453, 3143

\bibitem[{{Ghara} {$et~al$.}(2021{\natexlab{a}}){Ghara}, {Giri}, {Ciardi},
  {Mellema}, \& {Zaroubi}}]{2021MNRAS.503.4551G}
{Ghara}, R., {Giri}, S.~K., {Ciardi}, B., {Mellema}, G., \& {Zaroubi}, S.
  2021{\natexlab{a}}, \mnras, 503, 4551

\bibitem[{{Ghara} \& {Mellema}(2020)}]{ghara2020}
{Ghara}, R., \& {Mellema}, G. 2020, \mnras, 492, 634

\bibitem[{{Ghara} {$et~al$.}(2018){Ghara}, {Mellema}, {Giri}, {Choudhury},
  {Datta}, \& {Majumdar}}]{ghara18}
{Ghara}, R., {Mellema}, G., {Giri}, S.~K., {$et~al$.} 2018, \mnras, 476, 1741

\bibitem[{{Ghara} {$et~al$.}(2021{\natexlab{b}}){Ghara}, {Mellema}, \&
  {Zaroubi}}]{2021arXiv210813593G}
{Ghara}, R., {Mellema}, G., \& {Zaroubi}, S. 2021{\natexlab{b}}, arXiv
  e-prints, arXiv:2108.13593

\bibitem[{{Ghara} {$et~al$.}(2020){Ghara}, {Giri}, {Mellema}, {Ciardi},
  {Zaroubi}, {Iliev}, {Koopmans}, {Chapman}, {Gazagnes}, {Gehlot}, {Ghosh},
  {Jeli{\'c}}, {Mertens}, {Mondal}, {Schaye}, {Silva}, {Asad}, {Kooistra},
  {Mevius}, {Offringa}, {Pandey}, \& {Yatawatta}}]{2020MNRAS.493.4728G}
{Ghara}, R., {Giri}, S.~K., {Mellema}, G., {$et~al$.} 2020, \mnras, 493, 4728

\bibitem[{{Ghosh} {$et~al$.}(2012){Ghosh}, {Prasad}, {Bharadwaj}, {Ali}, \&
  {Chengalur}}]{Ghosh_2012}
{Ghosh}, A., {Prasad}, J., {Bharadwaj}, S., {Ali}, S.~S., \& {Chengalur}, J.~N.
  2012, \mnras, 426, 3295

\bibitem[{{Gillet} {$et~al$.}(2019){Gillet}, {Mesinger}, {Greig}, {Liu}, \&
  {Ucci}}]{2019MNRAS.484..282G}
{Gillet}, N., {Mesinger}, A., {Greig}, B., {Liu}, A., \& {Ucci}, G. 2019,
  \mnras, 484, 282

\bibitem[{{Giovannini}(2004)}]{Giovannini2004}
{Giovannini}, M. 2004, International Journal of Modern Physics D, 13, 391

\bibitem[{{Giri} {$et~al$.}(2019){Giri}, {D'Aloisio}, {Mellema}, {Komatsu},
  {Ghara}, \& {Majumdar}}]{2019JCAP...02..058G}
{Giri}, S.~K., {D'Aloisio}, A., {Mellema}, G., {$et~al$.} 2019, \jcap, 2019,
  058

\bibitem[{{Giri} \& {Mellema}(2021)}]{2021MNRAS.505.1863G}
{Giri}, S.~K., \& {Mellema}, G. 2021, \mnras, 505, 1863

\bibitem[{{Giri} {$et~al$.}(2018){Giri}, {Mellema}, \&
  {Ghara}}]{2018MNRAS.479.5596G}
{Giri}, S.~K., {Mellema}, G., \& {Ghara}, R. 2018, \mnras, 479, 5596

\bibitem[{{Glover} \& {Brand}(2003)}]{2003MNRAS.340..210G}
{Glover}, S.~C.~O., \& {Brand}, P.~W.~J.~L. 2003, \mnras, 340, 210

\bibitem[{{Grasso} \& {Rubinstein}(2001)}]{Grasso2001}
{Grasso}, D., \& {Rubinstein}, H.~R. 2001, \physrep, 348, 163

\bibitem[{{Greig} \& {Mesinger}(2017)}]{2017MNRAS.465.4838G}
{Greig}, B., \& {Mesinger}, A. 2017, \mnras, 465, 4838

\bibitem[{{Greig} \& {Mesinger}(2018)}]{2018MNRAS.477.3217G}
---. 2018, \mnras, 477, 3217

\bibitem[{{Greig} {$et~al$.}(2021{\natexlab{a}}){Greig}, {Mesinger},
  {Koopmans}, {Ciardi}, {Mellema}, {Zaroubi}, {Giri}, {Ghara}, {Ghosh},
  {Iliev}, {Mertens}, {Mondal}, {Offringa}, \& {Pandey}}]{2021MNRAS.501....1G}
{Greig}, B., {Mesinger}, A., {Koopmans}, L. V.~E., {$et~al$.}
  2021{\natexlab{a}}, \mnras, 501, 1

\bibitem[{{Greig} {$et~al$.}(2021{\natexlab{b}}){Greig}, {Mesinger},
  {Koopmans}, {Ciardi}, {Mellema}, {Zaroubi}, {Giri}, {Ghara}, {Ghosh},
  {Iliev}, {Mertens}, {Mondal}, {Offringa}, \& {Pandey}}]{2020arXiv200603203G}
---. 2021{\natexlab{b}}, \mnras, 501, 1

\bibitem[{{G{\"u}rkan} {$et~al$.}(2018){G{\"u}rkan}, {Hardcastle}, {Smith},
  {Best}, {Bourne}, {Calistro-Rivera}, {Heald}, {Jarvis}, {Prandoni},
  {R{\"o}ttgering}, {Sabater}, {Shimwell}, {Tasse}, \& {Williams}}]{gurkan2018}
{G{\"u}rkan}, G., {Hardcastle}, M.~J., {Smith}, D.~J.~B., {$et~al$.} 2018,
  \mnras, 475, 3010

\bibitem[{{Haiman}(2016)}]{2016ASSL..423....1H}
{Haiman}, Z. 2016, in Astrophysics and Space Science Library, Vol. 423,
  Understanding the Epoch of Cosmic Reionization: Challenges and Progress, ed.
  A.~{Mesinger}, 1

\bibitem[{{Harker} {$et~al$.}(2012){Harker}, {Pritchard}, {Burns}, \&
  {Bowman}}]{harker12}
{Harker}, G.~J.~A., {Pritchard}, J.~R., {Burns}, J.~O., \& {Bowman}, J.~D.
  2012, \mnras, 419, 1070

\bibitem[{{Hassan} {$et~al$.}(2017){Hassan}, {Dav{\'e}}, {Finlator}, \&
  {Santos}}]{2017MNRAS.468..122H}
{Hassan}, S., {Dav{\'e}}, R., {Finlator}, K., \& {Santos}, M.~G. 2017, \mnras,
  468, 122

\bibitem[{{Hibbard} {$et~al$.}(2022){Hibbard}, {Mirocha}, {Rapetti}, {Bassett},
  {Burns}, \& {Tauscher}}]{2022arXiv220102638H}
{Hibbard}, J.~J., {Mirocha}, J., {Rapetti}, D., {$et~al$.} 2022, arXiv
  e-prints, arXiv:2201.02638

\bibitem[{{Hills} {$et~al$.}(2018){Hills}, {Kulkarni}, {Meerburg}, \&
  {Puchwein}}]{Hills_2018}
{Hills}, R., {Kulkarni}, G., {Meerburg}, P.~D., \& {Puchwein}, E. 2018, \nat,
  564, E32

\bibitem[{{Iliev} {$et~al$.}(2012){Iliev}, {Mellema}, {Shapiro}, {Pen}, {Mao},
  {Koda}, \& {Ahn}}]{iliev12}
{Iliev}, I.~T., {Mellema}, G., {Shapiro}, P.~R., {$et~al$.} 2012, \mnras, 423,
  2222

\bibitem[{{Islam} {$et~al$.}(2019){Islam}, {Ghara}, {Paul}, {Choudhury}, \&
  {Nath}}]{2019MNRAS.487.2785I}
{Islam}, N., {Ghara}, R., {Paul}, B., {Choudhury}, T.~R., \& {Nath}, B.~B.
  2019, \mnras, 487, 2785

\bibitem[{{Jana} {$et~al$.}(2019){Jana}, {Nath}, \& {Biermann}}]{Jana_2019}
{Jana}, R., {Nath}, B.~B., \& {Biermann}, P.~L. 2019, \mnras, 483, 5329

\bibitem[{{Jensen} {$et~al$.}(2013){Jensen}, {Datta}, {Mellema}, {Chapman},
  {Abdalla}, {Iliev}, {Mao}, {Santos}, {Shapiro}, {Zaroubi}, {Bernardi},
  {Brentjens}, {de Bruyn}, {Ciardi}, {Harker}, {Jeli{\'c}}, {Kazemi},
  {Koopmans}, {Labropoulos}, {Martinez}, {Offringa}, {Pandey}, {Schaye},
  {Thomas}, {Veligatla}, {Vedantham}, \& {Yatawatta}}]{Jensen13}
{Jensen}, H., {Datta}, K.~K., {Mellema}, G., {$et~al$.} 2013, \mnras, 435, 460

\bibitem[{{Kamran} {$et~al$.}(2021{\natexlab{a}}){Kamran}, {Ghara}, {Majumdar},
  {Mondal}, {Mellema}, {Bharadwaj}, {Pritchard}, \&
  {Iliev}}]{2021MNRAS.502.3800K}
{Kamran}, M., {Ghara}, R., {Majumdar}, S., {$et~al$.} 2021{\natexlab{a}},
  \mnras, 502, 3800

\bibitem[{{Kamran} {$et~al$.}(2021{\natexlab{b}}){Kamran}, {Majumdar}, {Ghara},
  {Mellema}, {Bharadwaj}, {Pritchard}, {Mondal}, \&
  {Iliev}}]{2021arXiv210808201K}
{Kamran}, M., {Majumdar}, S., {Ghara}, R., {$et~al$.} 2021{\natexlab{b}}, arXiv
  e-prints, arXiv:2108.08201

\bibitem[{{Kang} \& {Jones}(2005)}]{2005ApJ...620...44K}
{Kang}, H., \& {Jones}, T.~W. 2005, \apj, 620, 44

\bibitem[{{Kapahtia} {$et~al$.}(2021){Kapahtia}, {Chingangbam}, {Ghara},
  {Appleby}, \& {Choudhury}}]{2021JCAP...05..026K}
{Kapahtia}, A., {Chingangbam}, P., {Ghara}, R., {Appleby}, S., \& {Choudhury},
  T.~R. 2021, \jcap, 2021, 026

\bibitem[{{Kern} {$et~al$.}(2017){Kern}, {Liu}, {Parsons}, {Mesinger}, \&
  {Greig}}]{2017ApJ...848...23K}
{Kern}, N.~S., {Liu}, A., {Parsons}, A.~R., {Mesinger}, A., \& {Greig}, B.
  2017, \apj, 848, 23

\bibitem[{{Kolopanis} {$et~al$.}(2019){Kolopanis}, {Jacobs}, {Cheng},
  {Parsons}, {Kohn}, {Pober}, {Aguirre}, {Ali}, {Bernardi}, {Bradley},
  {Carilli}, {DeBoer}, {Dexter}, {Dillon}, {Kerrigan}, {Klima}, {Liu},
  {MacMahon}, {Moore}, {Thyagarajan}, {Nunhokee}, {Walbrugh}, \&
  {Walker}}]{2019ApJ...883..133K}
{Kolopanis}, M., {Jacobs}, D.~C., {Cheng}, C., {$et~al$.} 2019, \apj, 883, 133

\bibitem[{{Kulsrud} \& {Pearce}(1969)}]{Kulsrud_1969}
{Kulsrud}, R., \& {Pearce}, W.~P. 1969, \apj, 156, 445

\bibitem[{{Kulsrud}(2004)}]{Kulsrud_2004}
{Kulsrud}, R.~M. 2004, {Plasma Physics for Astrophysics}

\bibitem[{{Kunze} \& {Komatsu}(2014)}]{KK14}
{Kunze}, K.~E., \& {Komatsu}, E. 2014, \jcap, 2014, 009

\bibitem[{{Lazar} \& {Bromm}(2022)}]{lazar2022}
{Lazar}, A., \& {Bromm}, V. 2022, \mnras, arXiv:2110.11956

\bibitem[{{Leite} {$et~al$.}(2017){Leite}, {Evoli}, {D'Angelo}, {Ciardi},
  {Sigl}, \& {Ferrara}}]{Leite_2017}
{Leite}, N., {Evoli}, C., {D'Angelo}, M., {$et~al$.} 2017, \mnras, 469, 416

\bibitem[{{Liu} \& {Parsons}(2016)}]{2016MNRAS.457.1864L}
{Liu}, A., \& {Parsons}, A.~R. 2016, \mnras, 457, 1864

\bibitem[{{Liu} {$et~al$.}(2019){Liu}, {Outmezguine}, {Redigolo}, \&
  {Volansky}}]{Liu2019prd}
{Liu}, H., {Outmezguine}, N.~J., {Redigolo}, D., \& {Volansky}, T. 2019, \prd,
  100, 123011

\bibitem[{Liu \& Slatyer(2018)}]{liu2018}
Liu, H., \& Slatyer, T.~R. 2018, Phys. Rev. D, 98, 023501

\bibitem[{{Madau} \& {Haardt}(2015)}]{2015ApJ...813L...8M}
{Madau}, P., \& {Haardt}, F. 2015, \apjl, 813, L8

\bibitem[{{Maity} \& {Choudhury}(2022)}]{2022MNRAS.511.2239M}
{Maity}, B., \& {Choudhury}, T.~R. 2022, \mnras, 511, 2239

\bibitem[{{Majumdar} {$et~al$.}(2013){Majumdar}, {Bharadwaj}, \&
  {Choudhury}}]{majumdar13}
{Majumdar}, S., {Bharadwaj}, S., \& {Choudhury}, T.~R. 2013, \mnras, 434, 1978

\bibitem[{{Majumdar} {$et~al$.}(2011){Majumdar}, {Bharadwaj}, {Datta}, \&
  {Choudhury}}]{majumdar11}
{Majumdar}, S., {Bharadwaj}, S., {Datta}, K.~K., \& {Choudhury}, T.~R. 2011,
  \mnras, 413, 1409

\bibitem[{{Majumdar} {$et~al$.}(2018){Majumdar}, {Pritchard}, {Mondal},
  {Watkinson}, {Bharadwaj}, \& {Mellema}}]{majumdar18}
{Majumdar}, S., {Pritchard}, J.~R., {Mondal}, R., {$et~al$.} 2018, \mnras, 476,
  4007

\bibitem[{{Mao} {$et~al$.}(2012){Mao}, {Shapiro}, {Mellema}, {Iliev}, {Koda},
  \& {Ahn}}]{mao12}
{Mao}, Y., {Shapiro}, P.~R., {Mellema}, G., {$et~al$.} 2012, \mnras, 422, 926

\bibitem[{{Mebane} {$et~al$.}(2018){Mebane}, {Mirocha}, \&
  {Furlanetto}}]{Mebane_2018}
{Mebane}, R.~H., {Mirocha}, J., \& {Furlanetto}, S.~R. 2018, \mnras, 479, 4544

\bibitem[{{Mellema} {$et~al$.}(2006){Mellema}, {Iliev}, {Pen}, \&
  {Shapiro}}]{mellema06}
{Mellema}, G., {Iliev}, I.~T., {Pen}, U.-L., \& {Shapiro}, P.~R. 2006, \mnras,
  372, 679

\bibitem[{{Mellema} {$et~al$.}(2015){Mellema}, {Koopmans}, {Shukla}, {Datta},
  {Mesinger}, \& {Majumdar}}]{2015aska.confE..10M}
{Mellema}, G., {Koopmans}, L., {Shukla}, H., {$et~al$.} 2015, Advancing
  Astrophysics with the Square Kilometre Array (AASKA14), 10

\bibitem[{{Mertens} {$et~al$.}(2020){Mertens}, {Mevius}, {Koopmans},
  {Offringa}, {Mellema}, {Zaroubi}, {Brentjens}, {Gan}, {Gehlot}, {Pandey},
  {Sardarabadi}, {Vedantham}, {Yatawatta}, {Asad}, {Ciardi}, {Chapman},
  {Gazagnes}, {Ghara}, {Ghosh}, {Giri}, {Iliev}, {Jeli{\'c}}, {Kooistra},
  {Mondal}, {Schaye}, \& {Silva}}]{martens20}
{Mertens}, F.~G., {Mevius}, M., {Koopmans}, L.~V.~E., {$et~al$.} 2020, \mnras,
  493, 1662

\bibitem[{{Mesinger} {$et~al$.}(2016){Mesinger}, {Greig}, \&
  {Sobacchi}}]{2016MNRAS.459.2342M}
{Mesinger}, A., {Greig}, B., \& {Sobacchi}, E. 2016, \mnras, 459, 2342

\bibitem[{{Mineo} {$et~al$.}(2012){Mineo}, {Gilfanov}, \&
  {Sunyaev}}]{2012MNRAS.419.2095M}
{Mineo}, S., {Gilfanov}, M., \& {Sunyaev}, R. 2012, \mnras, 419, 2095

\bibitem[{{Minoda} {$et~al$.}(2017){Minoda}, {Hasegawa}, {Tashiro}, {Ichiki},
  \& {Sugiyama}}]{Minoda17}
{Minoda}, T., {Hasegawa}, K., {Tashiro}, H., {Ichiki}, K., \& {Sugiyama}, N.
  2017, \prd, 96, 123525

\bibitem[{Minoda {$et~al$.}(2019)Minoda, Tashiro, \& Takahashi}]{Minoda19}
Minoda, T., Tashiro, H., \& Takahashi, T. 2019, Monthly Notices of the Royal
  Astronomical Society, 488, 2001

\bibitem[{{Mirocha} \& {Furlanetto}(2019{\natexlab{a}})}]{mirocha2019}
{Mirocha}, J., \& {Furlanetto}, S.~R. 2019{\natexlab{a}}, \mnras, 483, 1980

\bibitem[{{Mirocha} \& {Furlanetto}(2019{\natexlab{b}})}]{Mirocha_2019}
---. 2019{\natexlab{b}}, \mnras, 483, 1980

\bibitem[{{Mirocha} {$et~al$.}(2018){Mirocha}, {Mebane}, {Furlanetto},
  {Singal}, \& {Trinh}}]{2018MNRAS.478.5591M}
{Mirocha}, J., {Mebane}, R.~H., {Furlanetto}, S.~R., {Singal}, K., \& {Trinh},
  D. 2018, \mnras, 478, 5591

\bibitem[{{Mittal} \& {Kulkarni}(2021)}]{mittal21}
{Mittal}, S., \& {Kulkarni}, G. 2021, \mnras, 503, 4264

\bibitem[{{Mittal} \& {Kulkarni}(2022)}]{2022MNRAS.510.4992M}
---. 2022, \mnras, 510, 4992

\bibitem[{{Mondal} {$et~al$.}(2020){Mondal}, {Fialkov}, {Fling}, {Iliev},
  {Barkana}, {Ciardi}, {Mellema}, {Zaroubi}, {Koopmans}, {Mertens}, {Gehlot},
  {Ghara}, {Ghosh}, {Giri}, {Offringa}, \& {Pandey}}]{2020MNRAS.498.4178M}
{Mondal}, R., {Fialkov}, A., {Fling}, C., {$et~al$.} 2020, \mnras, 498, 4178

\bibitem[{{Mu{\~n}oz} {$et~al$.}(2015){Mu{\~n}oz}, {Kovetz}, \&
  {Ali-Ha{\"\i}moud}}]{Munoz15}
{Mu{\~n}oz}, J.~B., {Kovetz}, E.~D., \& {Ali-Ha{\"\i}moud}, Y. 2015, \prd, 92,
  083528

\bibitem[{{Mu{\~n}oz} \& {Loeb}(2018{\natexlab{a}})}]{Munoz18}
{Mu{\~n}oz}, J.~B., \& {Loeb}, A. 2018{\natexlab{a}}, \nat, 557, 684

\bibitem[{{Mu{\~n}oz} \& {Loeb}(2018{\natexlab{b}})}]{2018Natur.557..684M}
---. 2018{\natexlab{b}}, \nat, 557, 684

\bibitem[{{Mu{\~n}oz} {$et~al$.}(2022){Mu{\~n}oz}, {Qin}, {Mesinger}, {Murray},
  {Greig}, \& {Mason}}]{2022MNRAS.511.3657M}
{Mu{\~n}oz}, J.~B., {Qin}, Y., {Mesinger}, A., {$et~al$.} 2022, \mnras, 511,
  3657

\bibitem[{{Murmu} {$et~al$.}(2022){Murmu}, {Ghara}, {Majumdar}, \&
  {Datta}}]{Murmu2022joaa}
{Murmu}, C.~S., {Ghara}, R., {Majumdar}, S., \& {Datta}, K.~K. 2022, Journal of
  Astrophysics and Astronomy (accepted)

\bibitem[{{Natwariya} \& {Bhatt}(2020)}]{Natwariya2020}
{Natwariya}, P.~K., \& {Bhatt}, J.~R. 2020, arXiv e-prints, arXiv:2001.00194

\bibitem[{{Nebrin} {$et~al$.}(2019){Nebrin}, {Ghara}, \& {Mellema}}]{Nebrin19}
{Nebrin}, O., {Ghara}, R., \& {Mellema}, G. 2019, \jcap, 2019, 051

\bibitem[{{Paciga} {$et~al$.}(2013){Paciga}, {Albert}, {Bandura}, {Chang},
  {Gupta}, {Hirata}, {Odegova}, {Pen}, {Peterson}, {Roy}, {Shaw}, {Sigurdson},
  \& {Voytek}}]{paciga13}
{Paciga}, G., {Albert}, J.~G., {Bandura}, K., {$et~al$.} 2013, \mnras, 433, 639

\bibitem[{{Padmanabhan}(2021)}]{2021IJMPD..3030009P}
{Padmanabhan}, H. 2021, International Journal of Modern Physics D, 30, 2130009

\bibitem[{{Pal} {$et~al$.}(2021){Pal}, {Bharadwaj}, {Ghosh}, \&
  {Choudhuri}}]{Pal21}
{Pal}, S., {Bharadwaj}, S., {Ghosh}, A., \& {Choudhuri}, S. 2021, \mnras, 501,
  3378

\bibitem[{{Park} {$et~al$.}(2019){Park}, {Mesinger}, {Greig}, \&
  {Gillet}}]{2019MNRAS.484..933P}
{Park}, J., {Mesinger}, A., {Greig}, B., \& {Gillet}, N. 2019, \mnras, 484, 933

\bibitem[{{Parsons} {$et~al$.}(2021){Parsons}, {Mas-Ribas}, {Sun}, {Chang},
  {Gonzalez}, \& {Mebane}}]{parsons2021}
{Parsons}, J., {Mas-Ribas}, L., {Sun}, G., {$et~al$.} 2021, arXiv e-prints,
  arXiv:2112.06407

\bibitem[{{Partl} {$et~al$.}(2011){Partl}, {Maselli}, {Ciardi}, {Ferrara}, \&
  {M{\"u}ller}}]{2011MNRAS.414..428P}
{Partl}, A.~M., {Maselli}, A., {Ciardi}, B., {Ferrara}, A., \& {M{\"u}ller}, V.
  2011, \mnras, 414, 428

\bibitem[{{Patil} {$et~al$.}(2014){Patil}, {Zaroubi}, {Chapman}, {Jeli{\'c}},
  {Harker}, {Abdalla}, {Asad}, {Bernardi}, {Brentjens}, {de Bruyn}, {Bus},
  {Ciardi}, {Daiboo}, {Fernandez}, {Ghosh}, {Jensen}, {Kazemi}, {Koopmans},
  {Labropoulos}, {Mevius}, {Martinez}, {Mellema}, {Offringa}, {Pandey},
  {Schaye}, {Thomas}, {Vedantham}, {Veligatla}, {Wijnholds}, \&
  {Yatawatta}}]{patil2014}
{Patil}, A.~H., {Zaroubi}, S., {Chapman}, E., {$et~al$.} 2014, \mnras, 443,
  1113

\bibitem[{{Patil} {$et~al$.}(2017){Patil}, {Yatawatta}, {Koopmans}, {de Bruyn},
  {Brentjens}, {Zaroubi}, {Asad}, {Hatef}, {Jeli{\'c}}, {Mevius}, {Offringa},
  {Pandey}, {Vedantham}, {Abdalla}, {Brouw}, {Chapman}, {Ciardi}, {Gehlot},
  {Ghosh}, {Harker}, {Iliev}, {Kakiichi}, {Majumdar}, {Mellema}, {Silva},
  {Schaye}, {Vrbanec}, \& {Wijnholds}}]{2017ApJ...838...65P}
{Patil}, A.~H., {Yatawatta}, S., {Koopmans}, L.~V.~E., {$et~al$.} 2017, \apj,
  838, 65

\bibitem[{{Patwa} {$et~al$.}(2021){Patwa}, {Sethi}, \& {Dwarakanath}}]{patwa21}
{Patwa}, A.~K., {Sethi}, S., \& {Dwarakanath}, K.~S. 2021, \mnras, 504, 2062

\bibitem[{{Planck Collaboration} {$et~al$.}(2020){Planck Collaboration},
  {Aghanim}, {Akrami}, {Ashdown}, {Aumont}, {Baccigalupi}, {Ballardini},
  {Banday}, {Barreiro}, {Bartolo}, {Basak}, {Battye}, {Benabed}, {Bernard},
  {Bersanelli}, {Bielewicz}, {Bock}, {Bond}, {Borrill}, {Bouchet}, {Boulanger},
  {Bucher}, {Burigana}, {Butler}, {Calabrese}, {Cardoso}, {Carron},
  {Challinor}, {Chiang}, {Chluba}, {Colombo}, {Combet}, {Contreras}, {Crill},
  {Cuttaia}, {de Bernardis}, {de Zotti}, {Delabrouille}, {Delouis}, {Di
  Valentino}, {Diego}, {Dor{\'e}}, {Douspis}, {Ducout}, {Dupac}, {Dusini},
  {Efstathiou}, {Elsner}, {En{\ss}lin}, {Eriksen}, {Fantaye}, {Farhang},
  {Fergusson}, {Fernandez-Cobos}, {Finelli}, {Forastieri}, {Frailis},
  {Fraisse}, {Franceschi}, {Frolov}, {Galeotta}, {Galli}, {Ganga},
  {G{\'e}nova-Santos}, {Gerbino}, {Ghosh}, {Gonz{\'a}lez-Nuevo}, {G{\'o}rski},
  {Gratton}, {Gruppuso}, {Gudmundsson}, {Hamann}, {Handley}, {Hansen},
  {Herranz}, {Hildebrandt}, {Hivon}, {Huang}, {Jaffe}, {Jones}, {Karakci},
  {Keih{\"a}nen}, {Keskitalo}, {Kiiveri}, {Kim}, {Kisner}, {Knox},
  {Krachmalnicoff}, {Kunz}, {Kurki-Suonio}, {Lagache}, {Lamarre}, {Lasenby},
  {Lattanzi}, {Lawrence}, {Le Jeune}, {Lemos}, {Lesgourgues}, {Levrier},
  {Lewis}, {Liguori}, {Lilje}, {Lilley}, {Lindholm}, {L{\'o}pez-Caniego},
  {Lubin}, {Ma}, {Mac{\'\i}as-P{\'e}rez}, {Maggio}, {Maino}, {Mandolesi},
  {Mangilli}, {Marcos-Caballero}, {Maris}, {Martin}, {Martinelli},
  {Mart{\'\i}nez-Gonz{\'a}lez}, {Matarrese}, {Mauri}, {McEwen}, {Meinhold},
  {Melchiorri}, {Mennella}, {Migliaccio}, {Millea}, {Mitra},
  {Miville-Desch{\^e}nes}, {Molinari}, {Montier}, {Morgante}, {Moss}, {Natoli},
  {N{\o}rgaard-Nielsen}, {Pagano}, {Paoletti}, {Partridge}, {Patanchon},
  {Peiris}, {Perrotta}, {Pettorino}, {Piacentini}, {Polastri}, {Polenta},
  {Puget}, {Rachen}, {Reinecke}, {Remazeilles}, {Renzi}, {Rocha}, {Rosset},
  {Roudier}, {Rubi{\~n}o-Mart{\'\i}n}, {Ruiz-Granados}, {Salvati}, {Sandri},
  {Savelainen}, {Scott}, {Shellard}, {Sirignano}, {Sirri}, {Spencer},
  {Sunyaev}, {Suur-Uski}, {Tauber}, {Tavagnacco}, {Tenti}, {Toffolatti},
  {Tomasi}, {Trombetti}, {Valenziano}, {Valiviita}, {Van Tent}, {Vibert},
  {Vielva}, {Villa}, {Vittorio}, {Wandelt}, {Wehus}, {White}, {White},
  {Zacchei}, \& {Zonca}}]{Planck18}
{Planck Collaboration}, {Aghanim}, N., {Akrami}, Y., {$et~al$.} 2020, \aap,
  641, A6

\bibitem[{{Price} {$et~al$.}(2018){Price}, {Greenhill}, {Fialkov}, {Bernardi},
  {Garsden}, {Barsdell}, {Kocz}, {Anderson}, {Bourke}, {Craig}, {Dexter},
  {Dowell}, {Eastwood}, {Eftekhari}, {Ellingson}, {Hallinan}, {Hartman},
  {Kimberk}, {Lazio}, {Leiker}, {MacMahon}, {Monroe}, {Schinzel}, {Taylor},
  {Tong}, {Werthimer}, \& {Woody}}]{2018MNRAS.478.4193P}
{Price}, D.~C., {Greenhill}, L.~J., {Fialkov}, A., {$et~al$.} 2018, \mnras,
  478, 4193

\bibitem[{{Pritchard} \& {Furlanetto}(2006)}]{Pritchard_2006}
{Pritchard}, J.~R., \& {Furlanetto}, S.~R. 2006, \mnras, 367, 1057

\bibitem[{{Pritchard} \& {Furlanetto}(2007)}]{pritchard2007}
---. 2007, \mnras, 376, 1680

\bibitem[{{Pritchard} \& {Loeb}(2012)}]{pritchard12}
{Pritchard}, J.~R., \& {Loeb}, A. 2012, Reports on Progress in Physics, 75,
  086901

\bibitem[{{Ratra}(1992)}]{Ratra1992}
{Ratra}, B. 1992, \apjl, 391, L1

\bibitem[{{Reis} {$et~al$.}(2020){Reis}, {Fialkov}, \&
  {Barkana}}]{2020MNRAS.499.5993R}
{Reis}, I., {Fialkov}, A., \& {Barkana}, R. 2020, \mnras, 499, 5993

\bibitem[{{Ross} {$et~al$.}(2019){Ross}, {Dixon}, {Ghara}, {Iliev}, \&
  {Mellema}}]{2019MNRAS.487.1101R}
{Ross}, H.~E., {Dixon}, K.~L., {Ghara}, R., {Iliev}, I.~T., \& {Mellema}, G.
  2019, \mnras, 487, 1101

\bibitem[{{Ross} {$et~al$.}(2021){Ross}, {Giri}, {Mellema}, {Dixon}, {Ghara},
  \& {Iliev}}]{2021MNRAS.506.3717R}
{Ross}, H.~E., {Giri}, S.~K., {Mellema}, G., {$et~al$.} 2021, \mnras, 506, 3717

\bibitem[{{Samui}(2014)}]{Samui_2014}
{Samui}, S. 2014, \na, 30, 89

\bibitem[{{Sazonov} \& {Sunyaev}(2015)}]{Sazonov_2015}
{Sazonov}, S., \& {Sunyaev}, R. 2015, \mnras, 454, 3464

\bibitem[{{Schauer} {$et~al$.}(2019){Schauer}, {Liu}, \&
  {Bromm}}]{2019ApJ...877L...5S}
{Schauer}, A. T.~P., {Liu}, B., \& {Bromm}, V. 2019, \apjl, 877, L5

\bibitem[{{Schlickeiser}(2002)}]{schlickeiser2002cosmic}
{Schlickeiser}, R. 2002, {Cosmic Ray Astrophysics}

\bibitem[{{Schmit} \& {Pritchard}(2018)}]{2018MNRAS.475.1213S}
{Schmit}, C.~J., \& {Pritchard}, J.~R. 2018, \mnras, 475, 1213

\bibitem[{{Seager} {$et~al$.}(1999){Seager}, {Sasselov}, \&
  {Scott}}]{Seager1999}
{Seager}, S., {Sasselov}, D.~D., \& {Scott}, D. 1999, \apjl, 523, L1

\bibitem[{{Seager} {$et~al$.}(2000){Seager}, {Sasselov}, \&
  {Scott}}]{Seager2000}
---. 2000, \apjs, 128, 407

\bibitem[{{Seiffert} {$et~al$.}(2011){Seiffert}, {Fixsen}, {Kogut}, {Levin},
  {Limon}, {Lubin}, {Mirel}, {Singal}, {Villela}, {Wollack}, \&
  {Wuensche}}]{2011ApJ...734....6S}
{Seiffert}, M., {Fixsen}, D.~J., {Kogut}, A., {$et~al$.} 2011, \apj, 734, 6

\bibitem[{{Sethi}(2005)}]{sethi05}
{Sethi}, S.~K. 2005, \mnras, 363, 818

\bibitem[{{Sethi} \& {Subramanian}(2005)}]{SS05}
{Sethi}, S.~K., \& {Subramanian}, K. 2005, \mnras, 356, 778

\bibitem[{{Shaw} {$et~al$.}(2022){Shaw}, {Chakraborty}, {Kamran}, {Ghara},
  {Choudhuri}, {Ali}, {Pal}, A., {Kumar}, {Dutta}, \& {Sarkar}}]{skamo2}
{Shaw}, A.~K., {Chakraborty}, A., {Kamran}, M., {$et~al$.} 2022, (Manuscript in
  prep.)

\bibitem[{{Sims} \& {Pober}(2020)}]{Sims2020}
{Sims}, P.~H., \& {Pober}, J.~C. 2020, \mnras, 492, 22

\bibitem[{{Singh} \& {Subrahmanyan}(2019)}]{singh2019}
{Singh}, S., \& {Subrahmanyan}, R. 2019, \apj, 880, 26

\bibitem[{{Singh} {$et~al$.}(2021){Singh}, {Nambissan T.}, {Subrahmanyan},
  {Udaya Shankar}, {Girish}, {Raghunathan}, {Somashekar}, {Srivani}, \&
  {Sathyanarayana Rao}}]{Saurabh_2021}
{Singh}, S., {Nambissan T.}, J., {Subrahmanyan}, R., {$et~al$.} 2021, arXiv
  e-prints, arXiv:2112.06778

\bibitem[{{Sitwell} {$et~al$.}(2014){Sitwell}, {Mesinger}, {Ma}, \&
  {Sigurdson}}]{2014MNRAS.438.2664S}
{Sitwell}, M., {Mesinger}, A., {Ma}, Y.-Z., \& {Sigurdson}, K. 2014, \mnras,
  438, 2664

\bibitem[{{Skilling}(1975)}]{Skilling1975}
{Skilling}, J. 1975, \mnras, 172, 557

\bibitem[{{Tanaka} {$et~al$.}(2018){Tanaka}, {Hasegawa}, {Yajima}, {Kobayashi},
  \& {Sugiyama}}]{2018MNRAS.480.1925T}
{Tanaka}, T., {Hasegawa}, K., {Yajima}, H., {Kobayashi}, M. I.~N., \&
  {Sugiyama}, N. 2018, \mnras, 480, 1925

\bibitem[{{Tegmark} {$et~al$.}(1997){Tegmark}, {Silk}, {Rees}, {Blanchard},
  {Abel}, \& {Palla}}]{Tegmark_1997}
{Tegmark}, M., {Silk}, J., {Rees}, M.~J., {$et~al$.} 1997, \apj, 474, 1

\bibitem[{{Thomas} \& {Zaroubi}(2011)}]{thomas11}
{Thomas}, R.~M., \& {Zaroubi}, S. 2011, \mnras, 410, 1377

\bibitem[{Turner \& Widrow(1988)}]{PhysRevD.37.2743}
Turner, M.~S., \& Widrow, L.~M. 1988, Phys. Rev. D, 37, 2743

\bibitem[{{Venkatesan} {$et~al$.}(2001){Venkatesan}, {Giroux}, \&
  {Shull}}]{2001ApJ...563....1V}
{Venkatesan}, A., {Giroux}, M.~L., \& {Shull}, J.~M. 2001, \apj, 563, 1

\bibitem[{{Visbal} {$et~al$.}(2014){Visbal}, {Haiman}, {Terrazas}, {Bryan}, \&
  {Barkana}}]{visbal14}
{Visbal}, E., {Haiman}, Z., {Terrazas}, B., {Bryan}, G.~L., \& {Barkana}, R.
  2014, \mnras, 445, 107

\bibitem[{{Willott} {$et~al$.}(2010){Willott}, {Albert}, {Arzoumanian},
  {Bergeron}, {Crampton}, {Delorme}, {Hutchings}, {Omont}, {Reyl{\'e}}, \&
  {Schade}}]{2010AJ....140..546W}
{Willott}, C.~J., {Albert}, L., {Arzoumanian}, D., {$et~al$.} 2010, \aj, 140,
  546

\bibitem[{{Wouthuysen}(1952)}]{Wouthuysen_1952}
{Wouthuysen}, S.~A. 1952, \aj, 57, 31

\bibitem[{{Yajima} \& {Khochfar}(2015)}]{2015MNRAS.448..654Y}
{Yajima}, H., \& {Khochfar}, S. 2015, \mnras, 448, 654

\bibitem[{{Zaroubi} {$et~al$.}(2012){Zaroubi}, {de Bruyn}, {Harker}, {Thomas},
  {Labropolous}, {Jeli{\'c}}, {Koopmans}, {Brentjens}, {Bernardi}, {Ciardi},
  {Daiboo}, {Kazemi}, {Martinez-Rubi}, {Mellema}, {Offringa}, {Pandey},
  {Schaye}, {Veligatla}, {Vedantham}, \& {Yatawatta}}]{2012MNRAS.425.2964Z}
{Zaroubi}, S., {de Bruyn}, A.~G., {Harker}, G., {$et~al$.} 2012, \mnras, 425,
  2964

\end{thebibliography}

\end{document}